\documentclass[journal]{IEEEtran}
\usepackage{cite}
\usepackage{amsmath,amssymb,amsfonts}
\usepackage{graphicx}
\usepackage{textcomp}
\usepackage{subfigure}
\usepackage{adjustbox}
\usepackage{tabularx}
\usepackage{booktabs}
\usepackage{color}
\usepackage{multirow}
\usepackage{amsmath}

\usepackage{bm}

\usepackage[backref]{hyperref}
\usepackage{xparse}
\usepackage{multicol}
\usepackage{balance}
\usepackage{tabularx}

\usepackage{algorithm} 
\usepackage{tabularx}
\usepackage{hyperref}
\usepackage{multirow}
\usepackage{algpseudocode}  
\usepackage{amsmath}  
\usepackage{colortbl}
\definecolor{mygray}{gray}{0.8}
\ifCLASSINFOpdf
\else
\fi
\begin{document}
\title{Unsupervised Domain Adaptation for Low-dose CT Reconstruction via Bayesian Uncertainty Alignment }
\author{Kecheng Chen, Jie Liu, Renjie Wan, \IEEEmembership{Member, IEEE}, Victor Ho-Fun Lee, Varut Vardhanabhuti,
Hong Yan, \IEEEmembership{Fellow, IEEE}, and Haoliang Li, \IEEEmembership{Member, IEEE}
\thanks{This work is supported by the Hong Kong Innovation and Technology Commission (InnoHK Project ClMDA), the Hong Kong Research Grants Council (Projects 21200522 and 11204821), CityU New Research Initiatives/Infrastructure Support from Central (APRC
9610528), National Natural Science Foundation of China under Grant No. 62302415, Guangdong Basic and Applied Basic Research Foundation under Grant No. 2022A1515110692, 2024A1515012822, and the Blue Sky Research Fund of HKBU under Grant No. BSRF/21-22/16, and the Seed Fund for Collaborative Research (Project 2207101536), The University of Hong Kong. (Corresponding authors: Renjie Wan and Haoliang Li)}
\thanks{K. Chen, J. Liu, H. Yan, and H. Li are with the Department of Electrical Engineering and the Center for Intelligent Multidimensional Data Analysis, City University of Hong Kong, Hong Kong, China. (e-mail: cs.ckc96@gmail.com; jliu.ee@my.cityu.edu.hk; h.yan@cityu.edu.hk; haoliang.li@cityu.edu.hk.)}
\thanks{R. Wan is with the Department of Computer Science, Hong Kong Baptist University, Hong Kong, China. (e-mail: wanpeoplejie@gmail.com)}

\thanks{V. Lee is with the Department of Clinical Oncology, School of Clinical Medicine, LKS Faculty of Medicine, The University of Hong Kong, Hong Kong SAR, China. (e-mail: vhflee@hku.hk)}
\thanks{V. Vardhanabhuti is with the Department of Diagnostic Radiology, School of Clinical Medicine, LKS Faculty of Medicine, The University of Hong Kong, Hong Kong SAR, China. (e-mail: varv@hku.hk)}

}
% and the end of the author line. \textit{i.e.}, if you had this:

\markboth{Journal of \LaTeX\ Class Files,~Vol.~14, No.~8, August~2015}%
{Shell \MakeLowercase{\textit{et al.}}: Bare Demo of IEEEtran.cls for IEEE Journals}



\maketitle

\begin{abstract}
%%%
Low-dose computed tomography (LDCT) image reconstruction techniques can reduce patient radiation exposure while maintaining acceptable imaging quality. Deep learning is widely used in this problem, but the performance of testing data (\textit{a.k.a.} target domain) is often degraded in clinical scenarios due to the variations that were not encountered in training data (\textit{a.k.a.} source domain). Unsupervised domain adaptation (UDA) of LDCT reconstruction has been proposed to solve this problem through distribution alignment. However, existing UDA methods fail to explore the usage of uncertainty quantification, which is crucial for reliable intelligent medical systems in clinical scenarios with unexpected variations. Moreover, existing direct alignment for different patients would lead to content mismatch issues. To address these issues, we propose to leverage a probabilistic reconstruction framework to conduct a joint discrepancy minimization between source and target domains in both the latent and image spaces. In the latent space, we devise a Bayesian uncertainty alignment to reduce the epistemic gap between the two domains. This approach reduces the uncertainty level of target domain data, making it more likely to render well-reconstructed results on target domains. In the image space, we propose a sharpness-aware distribution alignment to achieve a match of second-order information, which can ensure that the reconstructed images from the target domain have similar sharpness to normal-dose CT  images from the source domain. Experimental results on two simulated datasets and one clinical low-dose imaging dataset show that our proposed method outperforms other methods in quantitative and visualized performance.

%%%%
\end{abstract}


%\linenumbers

\section{Introduction}
\label{intro}
\begin{figure*}[!t]
    \centering  
    \includegraphics[width=0.88\textwidth]{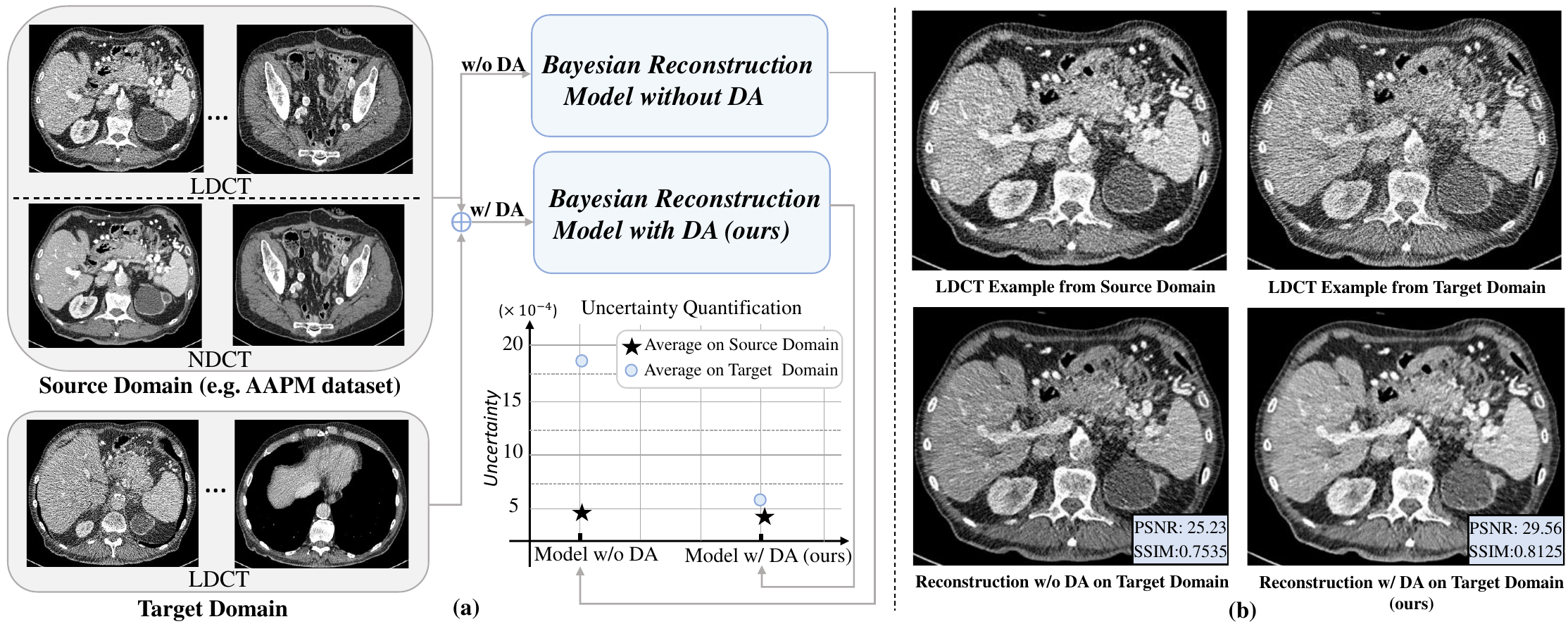}
    \caption{Left: Uncertainty quantification on source and target domains using BNN-based reconstruction model with DA strategy (\texttt{Model w/ DA} trained by paired data from the source domain and LDCT images from the target domain) and BNN-based reconstruction model without DA strategy (\texttt{Model w/o DA} trained by source domain data only). A higher level of uncertainty is observed on the target domain for the model without DA strategy. Besides, the level of uncertainty on the target domain is close to the source domain by adopting our proposed DA method. Right: Examples of an LDCT image from the source domain, an LDCT image from the target domain, and corresponding reconstructed results on target domain data. The display window is [-160,240] HU. } 
    \label{motivation}
\end{figure*}
Low-dose computed tomography (LDCT) image reconstruction techniques have the potential to decrease radiation exposure for patients while delivering images of nearly the same high quality as normal-dose CT (NDCT) images for radiologists \cite{9233366,paluru2021anam}. 
Among modern LDCT reconstruction approaches \cite{kulathilake2021review}, deep learning (DL)-based methods have become dominant in a post-processing manner compared with previous iterative reconstruction-based \cite{zeng2021deep,ikuta2022deep} and projection filtering-based methods \cite{ma2021sinogram}.  Thanks to the American Association of Physicists in Medicine (\texttt{AAPM}) that released simulated paired LDCT/NDCT benchmark datasets \cite{mccollough2020low}, quite a few DL models with elaborated network structures are proposed to learn a mapping from LDCT to NDCT images \cite{chen2021lesion}, achieving cutting-edge noise suppression and detail preservation. %LDCT image reconstruction techniques have the potential to decrease radiation exposure for patients while also delivering images of %nearly the same high quality as normal-dose CT (NDCT) images for radiologists.
However, when deploying trained models to real-world CT imaging scenarios, many variations in data acquisition that \textit{were not encountered during training} will be involved \cite{cobbinah2022reducing}, such as changes in software/hardware protocol from one institution to another (particularly reconstruction kernel), geometric factors (\textit{e.g.,} slice thickness and pixel spacing), and CT scanner manufacturer. For example, DL-based models trained on \texttt{AAPM} datasets or LDCT images from a particular scanner may not perform well in such diverse imaging situations, as these variations would violate the consistent assumption of feature or noise distribution between the training data (\textit{a.k.a.} source domain) and testing data (\textit{a.k.a.} target domain) \cite{li2023low,cobbinah2022reducing,liu2021coinet}. Note we focus on the LDCT post-reconstruction problem that aims to restore the LDCT images in the image domain without troublesomely accessing the projection data as discussed by \cite{yang2023transfer}.

% (or LDCT images with additional unpaired NDCT images)
To this end, it is necessary to address the domain adaptation (DA) problem of DL-based LDCT reconstruction. In most clinical scenarios, the LDCT images are available when patients were scanned by low-dose CT imaging mode. To reduce the noise interference of collected LDCT images, a plausible solution is to perform unsupervised learning (using LDCT images only) on the target domain directly. However, these unsupervised methods \cite{wu2019consensus,wang2023noise2noise} typically can be limited to their respective assumptions. For example, Wu \textit{et al.} \cite{wu2019consensus} utilized the Noise2Noise to LDCT images to reconstruct different noise realizations, 
but it may not be sufficient to remove the strong spatially corrected noise of LDCT images \cite{moran2020noisier2noise} due to its spatially uncorrected, zero-mean noise assumption. %The latter 
%will be limited if there are no unpaired NDCT images on the target domain, while more training time was reported due to the multiple adversarial losses for cycle consistency \cite{sharma2022comparative}. and an easier learning process \cite{daume2009frustratingly}
An alternative solution is to transfer the knowledge from the source domain to the target domain \cite{cai2019unsupervised}, as such, the model can be adaptive to the target domain well, leading to  better reconstruction performance \cite{lee2022unsupervised,li2023low,huang2023cddnet}. For this stream, unsupervised DA (UDA) methods adopt LDCT images from the target domain and easily accessible LDCT/NDCT image pairs (\textit{e.g.}, \texttt{AAPM} dataset) as the source domain -- making such UDA framework feasible in clinical scenarios. Recently, through the exploration of a similar NDCT-like style using \textit{conditional distribution alignment}, \textit{i.e.}, $p_{s}(\mathbf{y}|\mathbf{x}_{s}) \approx p_{t}(\mathbf{y}|\mathbf{x}_{t})$ (where $\mathbf{x}_{s}$, $\mathbf{x}_{t}$, $\mathbf{y}$ denote the input from the source and target domain, and the reconstruction result),
most UDA methods aim to generate indistinguishable reconstruction results for source and target domains by the guidance of adversarial domain classifiers \cite{lee2022unsupervised} or discrepancy minimization \cite{huang2023cddnet}.

%%%%%%%%%%%%%%%%%%%%%%% Version 1 %%%%%%%%%%%%%%%%%%%%%%%%%%%%%%%
%\jacky{However, one limitation is that the robustness of these UDA methods would be vulnerable for complex variations under cross-domain scenarios}, since they inherit deterministic models (\textit{e.g.}, convolutional layers) with \textit{fixed model weights} \cite{wilson2020bayesian}. Instead, the Bayesian neural network endows the distributional formula of model weights with better reconstruction robustness \cite{laves2022posterior,tolle2021mean,wilson2020bayesian}. More importantly, \jacky{the weight distribution is inherently capable of capturing the \textit{model uncertainty}\footnote{The model uncertainty can reflect models' epistemic gap between the test data and the training data \cite{kendall2017uncertainties}}}, which has been used to explore \jacky{out-of-distribution data with higher uncertainty} in visual \cite{nguyen2022out} and speech recognition communities \cite{oh2018modeling}. In the context of our cross-domain LDCT reconstruction, an obviously higher level of uncertainty can also be noticed (see Figure \hyperlink{motivation}{1a}{}, \texttt{Model w/o DA}) for target domain LDCT images, while unfavorable reconstruction results can be observed (see Figure \hyperlink{motivation}{1b}). Thus, the correlation  between model uncertainty and cross-domain performance motivates us to explore a potential uncertainty-guided
%method for the UDA problem of LDCT reconstruction.
%%%%%%%%%%%%%%%%%%%%%%% Version 2 %%%%%%%%%%%%%%%%%%%%%%%%%%%%%%%
Existing UDA methods, while effective, predominantly rely on deterministic models \cite{mallick2021deep}. However, these models lack the capability to quantify uncertainty in clinical scenarios \cite{laves2022posterior}, thereby limiting the deployment of intelligent medical systems.
Since the confidence of DL models' predictions would affect the diagnosis of physicians \cite{kompa2021second,liu2023robust,liu2024teller}, trustworthiness and reliability are crucial for medical systems.
Unlike deterministic models, the Bayesian neural network (BNN) with the distributional formula of model weights can capture the \textit{model uncertainty}\footnote{The model uncertainty can reflect models' epistemic gap between the test data and the training data \cite{kendall2017uncertainties}}, which has been used to detect out-of-distribution (OOD) data with higher uncertainty in wide applications \cite{nguyen2022out,oh2018modeling}. For cross-domain LDCT reconstruction, a correlation between model uncertainty and cross-domain performance can be observed (refer to Figure \hyperlink{motivation}{1a}), where target domain LDCT images exhibit higher uncertainty while displaying unfavorable reconstruction results (refer to Figure \hyperlink{motivation}{1b}). But it poses significant challenges to directly applying the BNN in the context of UDA LDCT reconstruction since it is unclear how to leverage explicit uncertainty for better reconstruction performance and how to ensure an effective uncertainty representation for the target domain without ground truth. Hence, we are dedicated to designing a BNN-based reconstruction framework with the uncertainty-guided method.

Another limitation of these methods arises from the style-oriented conditional distribution alignment \cite{huang2023cddnet,lee2022unsupervised}. In clinical scenarios, the LDCT images from the target domain and paired CT images from the source domain are usually collected from different patients. Consequently, the content information (\textit{i.e.}, anatomical region) of reconstruction images from two domains may be inconsistent in a batch (see Figure \hyperlink{motivation}{1a}). Directly applying style-oriented conditional distribution alignment to the reconstruction results of the two domains can cause potential confusion between style and content information, leading to difficulties in adversarial learning and potential mode collapse problems \cite{hassanpour2022survey}. It is, therefore,  necessary to alleviate the negative effect of \textit{content mismatch issue} and achieve a more effective conditional distribution alignment.
\begin{figure*}[!t]

    \centering
    \includegraphics[width=0.95\textwidth]{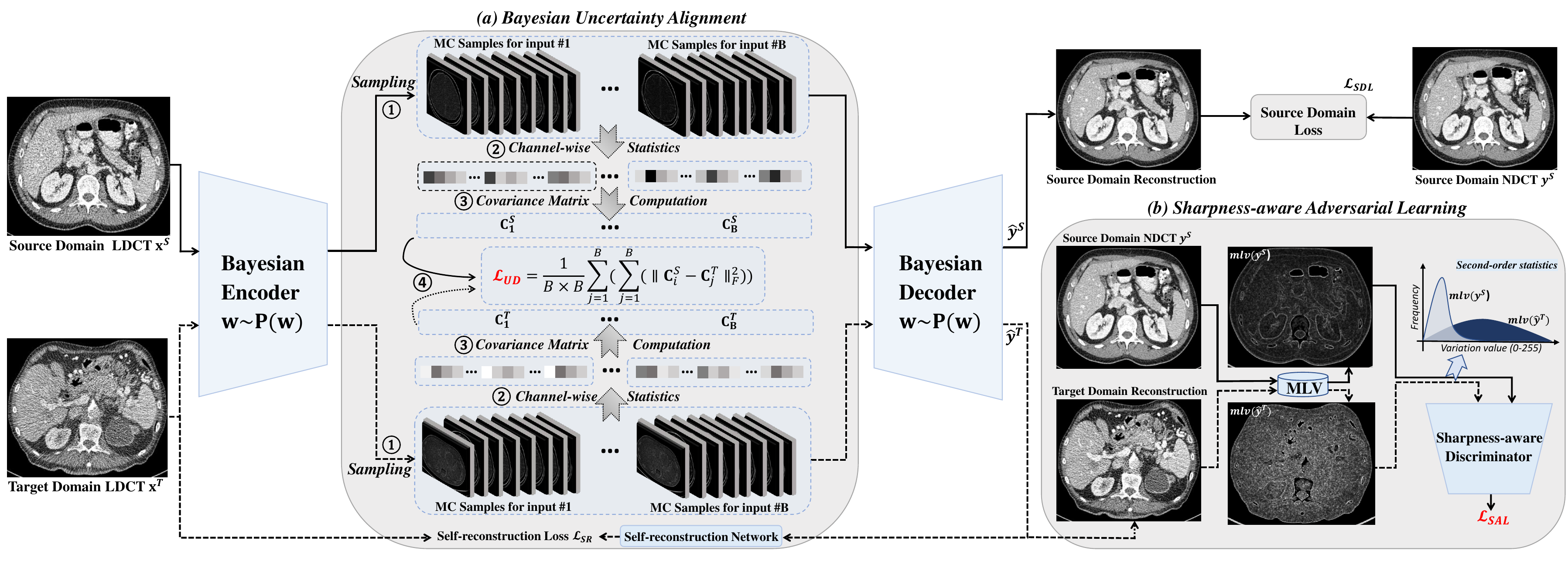}
    \caption{The overall framework of the proposed method. The CT reconstruction model is decoupled into a BNN-based encoder for the
feature extraction and a BNN-based decoder for the content reconstruction from extracted features. (a) \texttt{Bayesian uncertainty alignment} module (in the latent space): Uncertainty discrepancy term $\mathcal{L}_{UD}$ aims to explicitly quantify the uncertainty and reduce this discrepancy between source and target domains, as described in Section \ref{udml}; Self-reconstruction term $\mathcal{L}_{SR}$ as a complementary constraint encourages extracting
invertible latent features for the target domain in the process
of uncertainty discrepancy minimization, as described in section \ref{srl}. (b) \texttt{Sharpness-aware distribution alignment} module  (in the image space): This module encourages the model to reconstruct high-quality CT
images for target domains with a similar level of sharpness (as second-order information) as
the NDCT images from the source domain, as described in section \ref{Sharpness-aware Distribution}. }
    \label{framework}
\end{figure*}

%%%%%%%%%%%%%%%%%%%%%%% Version 1 %%%%%%%%%%%%%%%%%%%%%%%%%%%%%%%
%To address the abovementioned limitations, in this paper, a novel UDA method for cross-domain LDCT reconstruction is proposed, which incorporates Bayesian uncertainty-guided alignment and sharpness-aware distribution alignment using a BNN-based framework. As shown in Figure \ref{framework},  our whole framework can also be regarded as a joint discrepancy minimization between the source and target domains in the latent and image spaces. 
%Specifically, in the latent space, we  propose a novel Bayesian uncertainty alignment that directly reduces the gap in the level of uncertainty between the source and target domain data. Such uncertainty-guided alignment enforces the model to facilitate its epistemic ability for source-like target domain data, rendering accurate results on the target domain.
%Then, in the image space, we alleviate the \textit{content mismatch issue} by instead aligning the sharpness (\textit{i.e.} second-order information of image) between the reconstruction from the target domain and the NDCT from the source domain, given the observed sharpness differences between LDCT and NDCT images. To achieve this, we propose sharpness-aware adversarial learning by projecting reconstruction results into a generalized Gaussian distribution with decreased content interference using an effective sharpness descriptor.
%%%%%%%%%%%%%%%%%%%%%%% Version 2 %%%%%%%%%%%%%%%%%%%%%%%%%%%%%%%
To address the abovementioned limitations, in this paper, a novel UDA method for cross-domain LDCT reconstruction is proposed, which incorporates Bayesian uncertainty-guided alignment and sharpness-aware distribution alignment in the context of a BNN-based framework. As shown in Figure \ref{framework},  our whole framework can also be regarded as a joint discrepancy minimization between the source and target domains in the latent and image spaces. 
Specifically, in the latent space, we propose to directly reduce the gap in the level of uncertainty between the source and target domain data. To do so, we design an uncertainty discrepancy minimization loss based on the covariance matrix of Bayesian sampling embeddings to quantify the uncertainty gap better.
To ensure an effective uncertainty representation for the target
domain, we further introduce a simple self-reconstruction loss as a complementary constraint to encourage invertible latent features of the target domain in the process of uncertainty discrepancy minimization.
Such Bayesian uncertainty alignment can enforce the model to facilitate its epistemic ability for target domains similar to source domains, rendering accurate results on the target domain.
Then, in the image space, we alleviate the \textit{content mismatch issue} by instead aligning the sharpness, \textit{i.e.} second-order information of images, between the reconstruction from the target domain and the NDCT from the source domain, given the observed sharpness differences between LDCT and NDCT images. To achieve this, we propose sharpness-aware adversarial learning by projecting reconstruction results into a generalized Gaussian distribution with decreased content interference using an adequate sharpness descriptor.

%, but also relieve the mismatch effect of content information in such challenging alignment scenarios by the proxies of model uncertainty and the second-order statistics.

Our major contributions can be concluded as follows:
\begin{itemize}
    \item To the best of our knowledge, this is the first work that introduces the Bayesian neural network into the cross-domain LDCT reconstruction task.
    \item  We devise a Bayesian uncertainty alignment in the latent space to reduce the epistemic gap between source and target domains, making it more likely to render well-reconstructed results on the target domain.
    \item A sharpness-aware distribution alignment is proposed in the image space to reduce the interference of content mismatch in adversarial distribution alignment.
\end{itemize}

\section{Related works}
\subsection{Bayesian Neural Network}
Compared with deterministic models, probabilistic models such as Bayesian neural networks turn to find the maximum a posterior (MAP ) weights $\mathbf{w}^{\operatorname{MAP}}$ by placing a prior over the weights $\mathbf{w}$\cite{wang2020survey},
\begin{eqnarray}
\label{bayesian}
\mathbf{w}^{\operatorname{MAP}} &=& \arg\max_{\mathbf{w}}\log p(\mathbf{w}|D) \nonumber\\ &=&\arg\max_{\mathbf{w}}[\log p(D|\mathbf{w})+\log p(\mathbf{w})],
\label{map}
\end{eqnarray}
where $D$ denotes the training data. The first term is the complexity term and the second term is a prior distribution of the weight that can play a role of implicit regularization for better robustness \cite{mallick2021deep,wilson2020bayesian}.

As an important concept of the Bayesian neural network,  the \textit{Bayesian uncertainty} \cite{kendall2017uncertainties} can capture the model's ignorance about input data. For those out-of-domain (``unfamiliar") data, the model will lack sufficient epistemic capacities, leading to higher uncertainty \cite{nguyen2022out}. This model uncertainty can be estimated by the change in model parameters. For example, we can  calculate the variance of 
different predictions of the same input to do so \cite{hafner2018reliable} by using
Monte Carlo (MC) estimators with $T$ stochastic sampling operations from the weight distribution. For classification problems, the prediction uncertainty can be represented as follows \cite{cao2020uncertainty,feng2022bayesian}:
\begin{equation}
    \mathcal{U}_{var}(x) = \frac{1}{T}\sum_{t=1}^{T}(\hat{y}_{t}-\frac{1}{T}\sum_{t=1}^{T}\hat{y}_{t})^{2} \label{compute certainty}
\end{equation}
where each $\hat{y}_{t}$ is the prediction of the input $x$ based on sampled model weight $\mathbf{w}_{t}$ from $p(\mathbf{w})$.  Eq. \ref{compute certainty} can be used for regression problems to conduct pixel-wise uncertainty estimation easily, or can be rewritten as the form of the matrix computation \cite{kendall2017uncertainties}.

\subsection{Deep learning-based LDCT reconstruction}

Among various approaches, the filtering-based methods utilized classical filters (\textit{e.g.}, bilateral and statistical nonlinear filters \cite{sheikh2005visual}) to directly process the sinogram data of the CT image on the projection domain. Furthermore, the iterative reconstruction-based approaches \cite{yang2009linear,balda2012ray} conducted an iterative reconstruction (IR) between the projection and image domains. For example, Ikuta \textit{et al.} \cite{ikuta2022deep} proposed an IR formulation with a recurrent neural network to conduct dual-domain
learning between the projection and image domains. Although the abovementioned two streams of methods achieved acceptable performance, deploying these approaches in clinical scenarios is still challenging, due to the accessibility of the projection data and the time-consuming problem of the iterative reconstruction process \cite{chen2021lesion}. 

Post-processing-based LDCT reconstruction method \cite{johnson2016perceptual,zhang2023hierarchical} can enhance the quality of LDCT images on the image domain directly. We can first witness that some image priors (\textit{e.g.}, non-local similarity and total variation) are inserted into the ill-posed objective function to regularize the reconstruction process \cite{immonen2022use}, but, selecting the appropriate image priors manually can be a challenging task. Instead, deep learning (DL)-based LDCT reconstruction methods automatically learn implicit image priors from LDCT images \cite{fan2019quadratic}.

Recently, the domain adaptation problem of medical image reconstruction, e.g., LDCT image and magnetic resonance imaging (MRI) data \cite{tan2022fourier}, becomes very emergent due to its clinical practice \cite{lee2022unsupervised,huang2023cddnet}.  Among these approaches, GAN-NETL \cite{li2023low}  utilized paired Phantom data acquired from the target domain beforehand to fine-tune the pre-trained model of the source domain. However, collecting paired data (even if few Phantom data) from the target domain is extremely expensive by such a supervised domain adaptation method. Instead,  Kang \textit{et al.} \cite{kang2019cycle} used CycleGAN algorithm to learn an image-to-image translation by unpaired LDCT/NDCT images. Nevertheless, such a ClycleGAN-based scheme may suffer from potential mode collapse issues when there is no good feature discrimination between domains \cite{hassanpour2022survey}. More importantly, complex noise with diverse variations in LDCT images will further impede an effective style transfer. Finally, extra data requisition on target domains, \textit{i.e.,} unpaired NDCT images, may reduce the flexibility of this method.  As a result, we aim to develop an easy yet effective unsupervised domain adaptation (UDA) framework for the LDCT reconstruction problem.

\section{Methodology}
\subsection{Problem Statement}
In this paper, we address the unsupervised domain adaptation (UDA) problem for LDCT reconstruction. Assume that pairs of LDCT and NDCT images as the source domain can be collected from a public simulated benchmark dataset (\textit{e.g.}, \texttt{AAPM} dataset) or a particular scanner, represented as $D_{S} = \{(\mathbf{x}_{1}^{S},\mathbf{y}_{1}^{S}),\cdots, (\mathbf{x}_{N}^{S},\mathbf{y}_{N}^{S}))\}$ where $N$
denotes the number of pairs. Note that LDCT and NDCT images in the source domain strictly correspond for each pair of slices, which can ensure a reliable and correct source domain-related supervised loss (will be described later).
%The testing LDCT images (represented as $D_{T} = \{\mathbf{x}_{1}^{T},\cdots, \mathbf{x}_{L}^T\}$, where $L$ denotes the number of LDCT images) as the target domain may be collected in a situation different from source domains, such as changes in software/hardware protocol from one institution to another (particularly reconstruction kernel), geometric factors (\textit{e.g.}, slice thickness and pixel spacing), and CT scanner manufacturer. 
The testing LDCT images in the target domain are represented as $D_{T} = \{\mathbf{x}_{1}^{T},\cdots, \mathbf{x}_{L}^T\}$, where $L$ denotes the number of LDCT images. These images have been collected under different circumstances than the source domain, such as changes in software/hardware protocols from one institution to another (particularly reconstruction kernel), geometric factors (\textit{e.g.}, slice thickness and pixel spacing), and CT scanner manufacturer.
Such complex variations would cause a potential distribution inconsistency (a.k.a. domain shift) problem between source and target domains. By using paired LDCT/NDCT images on $D_{S}$ and LDCT images on $D_{T}$, our objective is to learn a model $F$ that can adapt well to the target domain $D_{T}$, resulting in high-quality reconstructed NDCT images $\mathbf{\hat{y}}^{T}= F(\mathbf{x^{T}})$.
%our objective is to learn a model $F$, which can be adaptive well on the target domain $D_{T}$ such that the high-quality reconstructed NDCT image $\mathbf{\hat{y}}^{T}= F(\mathbf{x^{T}})$ on the target domain can be obtained. 

%through alleviating the noise distribution shift between $D_{S}$ and $D_{T}$. Finally, 

%Overall,  we propose to construct a robust cross-domain CT reconstruction model under a probabilistic framework (section B). Based on this Bayesian-endowed CT reconstruction framework, we propose to alleviate noise distribution shifts between source and target domains from two perspectives (\textit{i.e.}, feature and image spaces). First, a 
%novel implicit noise distribution alignment method is proposed by means of the Bayesian uncertainty in the feature space (section C). Second, we propose to align the residual information (as the approximation of noise distribution)  between two domains via an adversarial manner in image space.

\subsection{LDCT reconstruction under a probabilistic framework}
In this paper, instead of using deterministic models (\textit{e.g.}, convolutional autoencoders) as the backbone reconstruction network, we introduce a probabilistic autoencoder based on the Bayesian neural network (BNN) due to its better robustness and uncertainty quantification \cite{laves2022posterior,tolle2021mean,wilson2020bayesian}. 

As shown in Figure \ref{framework}, we propose to disentangle such a reconstruction model into two decoupled probabilistic components, \textit{i.e.}, $F = G \circ E $. Specifically, a BNN-based encoder $E(\cdot)$ can project the input into the latent space for the feature extraction. Besides, 
a BNN-based decoder $G(\cdot)$ can reconstruct extracted features to the image space with high-quality CT images. For the distribution formula of the weight in the BNN, it is difficult to infer an exact solution, as the computation of posterior distribution over the weight $\mathbf{w}$, \textit{i.e.}, $ p(\mathbf{w}|\mathbf{x},\mathbf{y})=p(\mathbf{y}|\mathbf{w},\mathbf{x})p(\mathbf{w})/p(\mathbf{y}|\mathbf{x})$ is analytically intractable. Thus, we adopt an approximated posterior distribution over the weight via variational inference, \textit{i.e.},
\begin{equation}
\label{variational_inference}
    \theta^{*} = \arg \min_{\theta} \operatorname{KL}[q(\mathbf{w}|\theta)||p(\mathbf{w}|\mathbf{x},\mathbf{y})],
\end{equation}
where $q(\cdot)$ is a variational distribution represented by the distributional parameters $\theta$. Our goal is to find appropriate parameters $\theta^{*}$ of variational distribution such that the Kullback-Leibler (KL) divergence between the true posterior distribution over the weight $p(\mathbf{w}|\mathbf{x},\mathbf{y})$ and the variational distribution $q(\mathbf{w}|\theta)$ is minimized. Furthermore, Eq. \ref{variational_inference} can be reformulated as,
\begin{eqnarray}
    \theta^{*} &=&\arg\min_{\theta}\operatorname{KL[q(\mathbf{w}|\theta)||p(\mathbf{w})]} \nonumber \\ &-&\mathbb{E}_{\mathbf{w}\sim q(\mathbf{w}|\theta)}[\log p(\mathbf{x},\mathbf{y}|\mathbf{w})],
\end{eqnarray}
where the first term can balance the model complexity with its prior distribution $p(\mathbf{w})$. Moreover, the second term acts in an analogous way to the first term in Eq. (\ref{map}), which aims to fit the training data well as the same as deterministic models (as a reconstruction loss). This objective is so-called evidence lower bound (ELBO) loss. %Note that the advantage results from the first term as a complexity term, which can render the balance between the data fitting and model complexity.  (\textit{\textit{e.g.}}, Mean Absolute Error (MAE))
By incorporating the re-parameter trick \cite{tian2020learning} and stochastic gradient descent method, the variational parameter $\theta$ can be solved flexibly.

Based on this decoupled probabilistic LDCT reconstruction framework,  we will describe how to perform a joint discrepancy minimization between the source and target domains in the latent (\textit{i.e.}, the output of encoder $E(\cdot)$)
and image (\textit{i.e.}, the output of decoder $G(\cdot)$) spaces.

%This scheme will also contribute to the subsequent operations
%for addressing noise distribution shifts problem
%for the alleviation of noise distribution shifts, as %alleviating the noise distribution shift between source and %target domains can further render more robust CT %reconstruction performance on the target domain compared with %the gain of the introduction of probabilistic framework. 

%needs to explored in the following sections.
%Although the introduction of the probabilistic framework is helpful for the robust cross-domain CT reconstruction, the noise distribution shift between source and target domains need to be alleviated such that 
%To derive a robust cross-domain CT reconstruction model, an intuitive idea is to learn a domain-invariant encoder and decoder, respectively. By doing so, better and robust reconstruction performance can be achieved on target domain.

%learn a more generalizable model that can tackle the cross-domain robust CT reconstruction problem in the following sections.

\subsection{Domain Gap Minimization by Bayesian Uncertainty Alignment}
As discussed in the section \ref{intro}, the trained Bayesian model will lack sufficient epistemic ability for OOD)data (\textit{e.g.}, the target domain data in the DA problem), leading to higher uncertainty. Our empirical observations in Figure \hyperlink{motivation}{1a} further coincide with the higher uncertainty on the target domain with unfavorable results in cross-domain LDCT reconstruction problem. In light of this, we propose a Bayesian uncertainty alignment method to directly reduce the gap in the level of uncertainty between the source and target domains, making it more likely to render well-reconstructed results on the target domain. Due to more abstract representations of latent features as well as computational efficiency, recent OOD detection works \cite{nguyen2022out} had shown better performance based on the uncertainty over latent features instead of final predictions. Meanwhile, our preliminary experiments found negligible gain based on the final reconstruction results. Thus, the proposed Bayesian uncertainty alignment method is constructed in the latent space. 

However, achieving such uncertainty-guided alignment in the latent space will encounter two potential problems, as follows:
\begin{itemize}
    \item \textit{Uncertainty quantification.} Existing OOD detection methods implicitly leverage the uncertainty over latent features to generate randomized embeddings \cite{nguyen2022out} by multiple MC sampling. It is still unclear how to explicitly quantify the uncertainty over latent features with multiple feature maps in the cross-domain LDCT reconstruction problem.
    \item \textit{Effective representation for the target domain.} Although the encoder shared with paired source domain data may be used for latent feature representation for target domain data, it does not fully guarantee whether latent representations for the target domain are invertible and effective due to the unsupervised property of target domain data.
\end{itemize}

To address these problems, as shown in Figure \hyperlink{framwork}{2a}, we first propose an uncertainty discrepancy minimization loss, which leverages the covariance matrix of latent features to quantify the uncertainty among features. Moreover, we introduce a self-reconstruction loss based on self-supervised information to encourage invertible and effective latent representations for the target domain. More details will be described as follows:

\subsubsection{Uncertainty discrepancy minimization loss} \label{udml} Specifically, by feeding two batches of LDCT samples from source and target domains, \textit{i.e.}, $\{\mathbf{x}_{i}^{S}\}_{i=1}^{B}$ and $\{\mathbf{x}_{i}^{T}\}_{i=1}^{B}$ (where $\mathbf{x}_{i}^{S(T)} \in \mathbb{R}^{W\times H \times 1}$ denotes a LDCT sample with spatial size of $W\times H$ and the channel number of 1. $B$ denotes the number of images in a mini-batch), into a shared probabilistic encoder $E$, the output is a probabilistic embedding for each sample, \textit{i.e.}, $P(\mathbf{z}_{i}^{S(T)}|\mathbf{x}_{i}^{S(T)}) = P(\mathbf{z}_{i}^{S(T)}|\mathbf{x}_{i}^{S(T)},\mathbf{w}_{E})$, where $\mathbf{w}_{E}\sim P(\mathbf{w}_{E})$. The predictive distribution
of $P(\mathbf{z}^{S(T)}|\mathbf{x}^{S(T)})$ can be unbiased approximation using Monte
Carlo (MC) estimators with $M$ stochastic sampling operations over the $\mathbf{w}_{E}$, \textit{i.e.},
\begin{equation}
\label{mc}
    \mathbf{z}_{i}^{S(T)} = \mathbb{E}[P(\mathbf{z}_{i}^{S(T)}|\mathbf{x}_{i}^{S(T)},\mathbf{w}_{E})] = \frac{1}{M}\sum_{j=1}^{M}E_{\mathbf{w}_{E}^{j}}(\mathbf{x}_{i}^{S(T)}),
\end{equation}
where $E_{\mathbf{w}_{E}}(\cdot)$ denotes the parameterized probabilistic encoder, and $\mathbf{z}_{i}^{S(T)} \in \mathbb{R}^{W^{'}\times H^{'}\times C}$ with $C$ feature channels.
By imposing the virtue of probabilistic framework in Eq. \ref{mc}, we can derive different Bayesian sampling embeddings of the same input $\mathbf{x}_{i}$ to render Bayesian uncertainty estimation, \textit{i.e.},$
   \mathbf{Z}_{i}^{S(T)} = \{\mathbf{z}_{i,j}^{S(T)}\}_{j=1}^{M}$,
where the subscript $j$ denotes the index of MC sampling and 
$\mathbf{Z}_{i}^{S(T)} \in \mathbb{R}^{M\times W^{'}\times H^{'}\times C}$. Note that the outputs of the last layer in the encoder are only adopted as the probabilistic embeddings, as these latent features of this layer are more abstract compared with that of  previous layers.

Here, it is necessary to attain a more compact latent feature representation for Bayesian uncertainty estimation, as the information on spatial dimensions $H^{'}\times W^{'}$ are redundant as discussed by \cite{hu2018squeeze}. Usually, the channel-wise statistics can reflect more expressive and representative latent features by shrinking the spatial dimensions of latent embeddings, which is observed by previous lectures \cite{yang2009linear,shen2015multi}. 

By doing so, as shown in Figure \hyperlink{framework}{1a}, we utilize the squeeze module in \cite{hu2018squeeze} to conduct global information embedding via global average pooling. As such, we can derive more compact and representative latent features for Bayesian uncertainty estimation in latent space. Specifically, the global average pooling can be imposed for each channel element of each sampling embedding, \textit{i.e.}, $\mathbf{z'}^{S(T)} \in \mathbb{R}^{W^{'}\times H^{'}}$ as follows:
\begin{eqnarray}
\label{se}
  u^{S}= F_{sq}(\mathbf{z'}^{S}) = \frac{1}{W^{'}\times H^{'}}\sum_{t=1}^{W^{'}}\sum_{q=1}^{H^{'}}z_{t,q}^{S}, \nonumber \\
  u^{T}= F_{sq}(\mathbf{z'}^{T}) = \frac{1}{W^{'}\times H^{'}}\sum_{t=1}^{W^{'}}\sum_{q=1}^{H^{'}}z_{t,q}^{T},
\end{eqnarray}
where $F_{sq}(\cdot)$ denotes the global average pooling. Finally, the compact latent feature representation can be calculated as $\mathbf{U}_{i}^{S(T)} = F_{sq}(\mathbf{Z}_{i}^{S(T)})$, where $\mathbf{U}_{i}^{S(T)} \in \mathbb{R}^{M\times C}$.  More importantly, we aim to minimize the Bayesian uncertainty discrepancy of obtained latent features $\mathbf{U}_{i}^{S(T)}$  between source and target domains. Thus, an explicit uncertainty quantification is necessary for each domain. 

To this end, we propose to leverage the covariance matrix of latent features, as second-order statistics of latent features, to quantify the uncertainty among features. % as an implicit representation of the noise distribution. 
 Specifically, we propose to calculate the covariance matrix over different Bayesian sampling embeddings of the same input, \textit{i.e.},$\mathbf{U}_{i}^{S(T)}$ as follows:

\begin{eqnarray}
\label{covariance}
    \mathbf{C}_{i}^{S} &=&\frac{1}{M-1}\sum_{j=1}^{M} (\mathbf{u}_{i,j}^{S}-\bm{\mu}_{i}^{S})^{t}(\mathbf{u}_{i,j}^{S}-\bm{\mu}_{i}^{S}), \nonumber \\
    \mathbf{C}_{i}^{T} &=&\frac{1}{M-1}\sum_{j=1}^{M} (\mathbf{u}_{i,j}^{T}-\bm{\mu}_{i}^{T})^{t}(\mathbf{u}_{i,j}^{T}-\bm{\mu}_{i}^{T}),
\end{eqnarray}
where $\bm{\mu}_{i}^{S}$ and $\bm{\mu}_{i}^{T}$ denote the mean of the $\mathbf{U}_{i}^{S}$ and $\mathbf{U}_{i}^{T}$, respectively. $\mathbf{u}_{i,j}^{S}$ and  $\mathbf{u}_{i,j}^{T}$ represent the $j$-th row of the $\mathbf{U}_{i}^{S}$ and $\mathbf{U}_{i}^{T}$. The size of  $\mathbf{C}_{i}^{S}$ and $\mathbf{C}_{i}^{T}$ is $C \times C$. 

Note that the calculation in Eq. \ref{covariance} is different from the existing covariance matrix alignment of latent features (\textit{e.g.}, CORAL\cite{sun2016return}) in cross-domain classification problems as the covariance matrix in Eq. \ref{covariance} is derived from the different Bayesian sampling embeddings of a same sample rather than different samples in the source (or target) domain. The rationality of the covariance matrix for uncertainty quantification is that our proposed method can ignore the interference of content information (as the content information are same for different sampling embeddings of the same input), thereby the variations of each same dimension (\textit{i.e.}, uncertainty information) in latent features can be concentrated furthermore.

Finally, the uncertainty discrepancy as a $\mathcal{L}_{UD}$ objective in a batch of samples between source and target domains can be minimized as below,
\begin{equation}
\label{bnua}
    \mathcal{L}_{UD} = \frac{1}{B\times B}\sum_{i=1}^{B}\sum_{j=1}^{B}\| \mathbf{C}_{i}^{S} - \mathbf{C}_{j}^{T}\|_{F}^{2},
\end{equation}
where $\|\cdot\|_{F}$ denotes the Frobenius norm.

\subsubsection{Self-recontruction loss} \label{srl}In the process of uncertainty discrepancy minimization, it is also necessary to guarantee an effective and invertible latent representation of the target domain.  Due to the lack of supervised information on the target domain, we instead utilize the self-supervised information to do so. Specifically,  a simple self-reconstruction network is constructed to inverse reconstruction results $\hat{\mathbf{y}}^{T}$ of the target domain to the original LDCT images. Such strategy has demonstrated effectiveness in previous image enhancement-related works \cite{wan2022purifying}. The self-reconstruction loss can be represented as follows:
\begin{equation}
    \mathcal{L}_{SR} = \|SR(\hat{\mathbf{y}}^{T})-\mathbf{x}^{T}\|_{1}, \label{self}
\end{equation}
where $\hat{\mathbf{y}}^{T}$=$G(E(\mathbf{x}^{T}))$ and $SR(\cdot)$ denotes a self-reconstruction network.  Eq. \ref{self} evaluates whether the reconstructed result $\hat{\mathbf{y}}$ of the target domain can accurately return to the original input space. Thus, such self-supervised information can encourage the shared encoder to project effective and invertible latent features for the target domain. 

Finally, the proposed Bayesian uncertainty alignment method consists of two complementary terms. \textit{i.e.},
\begin{equation}
    \mathcal{L}_{BUA} =  \mathcal{L}_{UD} + \mathcal{L}_{SR},
\end{equation}
where the first term aims to reduce the epistemic gap between source and target domains in the latent space through uncertainty discrepancy. Meanwhile, to ensure the effectiveness of latent representations for the target domain, the second term as a complementary constraint encourages extracting invertible latent features for the target domain in the process of uncertainty discrepancy minimization.

\textbf{Discussion.} Our  proposed Bayesian uncertainty alignment method has several advantages, as follows: First, the computation of Eq. \ref{bnua} does not refer to any NDCT images on the target domain $D_{T}$, which is an unsupervised manner. Second, compared with existing UDA methods for cross-domain LDCT reconstruction, the proposed Bayesian uncertainty alignment module introduces an additional discrepancy minimization in the latent space, which uses the model uncertainty as a proxy to reduce the epistemic  capacity for target domain data.
    \begin{figure}[!t]
        \centering
        \label{mlv_figre}
        \includegraphics[width=\linewidth]{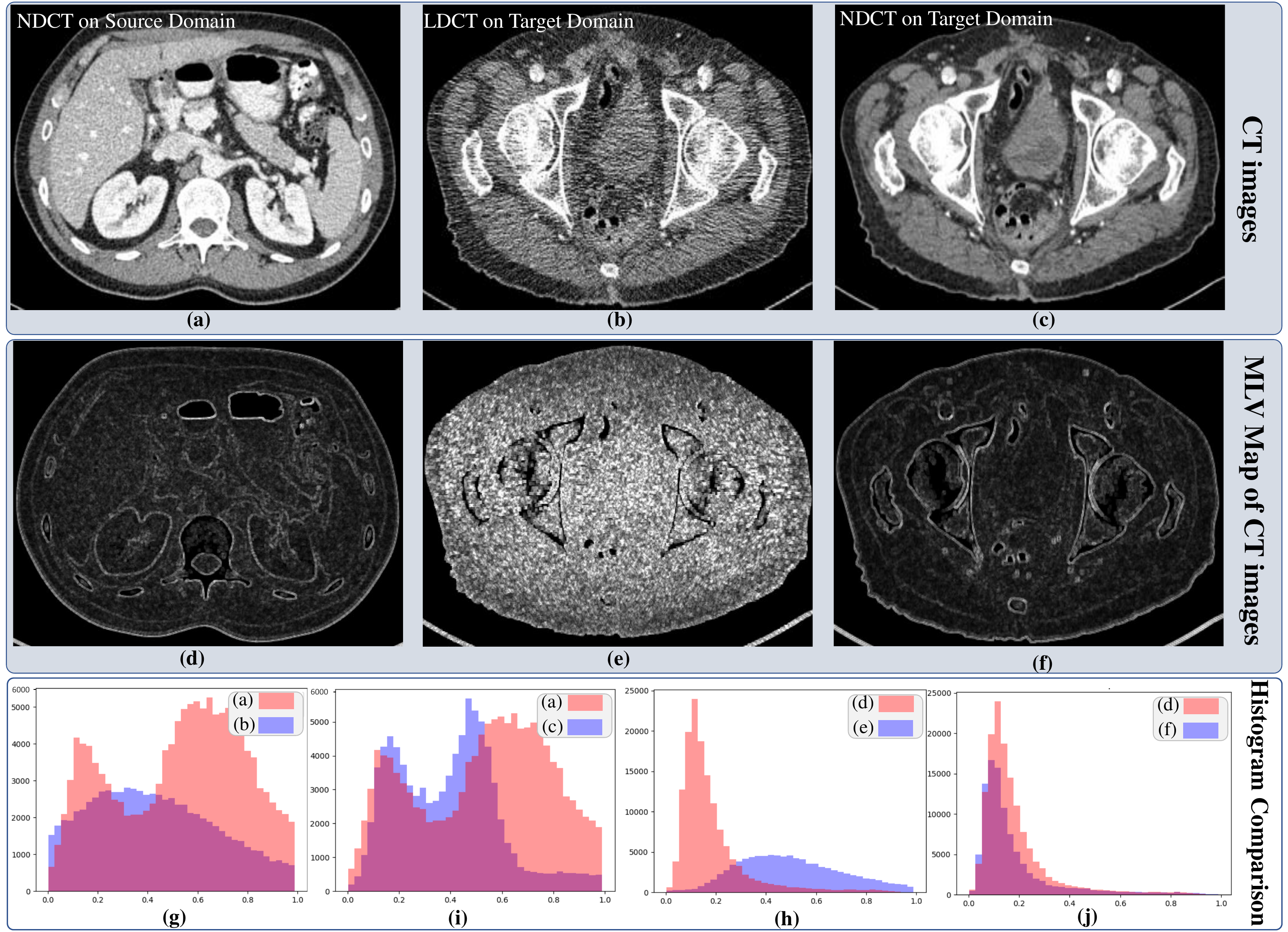}
        \caption{\label{fig:LFA} (a): An example NDCT from source domain; (b) An example LDCT from target domain; (c) NDCT version of (b); (d), (e) and (f) is the visualized results of MLV map for (a), (b) and (c). (g), (i) (h) and (j) are a series of histogram comparisons of different images in the first two rows.  Comparison between (g) and (i) implies that there is still an obvious distribution discrepancy even if the LDCT images (b) on the target domain have been reconstructed well on (c), due to a mixture of content and style information. Instead,  (h) implies that the sharpness-based histogram distribution is well separable. (j) show a high distribution consistency if the LDCT images (b) on the target domain have been reconstructed well. }
    \end{figure}
\subsection{Sharpness-aware Distribution Alignment via Adversarial Training}
\label{Sharpness-aware Distribution}
In the previous section, we figure out how to conduct Bayesian uncertainty alignment in the latent space.  Meanwhile, it is also important 
to reduce the discrepancy of conditional distribution (\textit{i.e.}, $p_{s}(\mathbf{y}_{s}|\mathbf{x}_{s})$ \textit{v.s.} $p_{t}(\mathbf{y}_{t}|\mathbf{x}_{t})$) between source and target domains in the image space, such that high-quality reconstruction results with NDCT-like style can be achieved. In this section, we will describe how to alleviate the mismatch problem of content information in the style-oriented conditional distribution alignment. 

Specifically, instead of  using reconstruction results directly, we propose to utilize the second-order information of reconstruction results, \textit{i.e.}, the sharpness,  to conduct conditional distribution alignment. %Bewe lack an extra regularization on the reconstructed CT images in the image space. Obviously, we hope that the 
Here, the maximum local variation (MLV) \cite{bahrami2014fast} as a sharpness descriptor $\Phi(\cdot)$ is introduced to generate the MLV map for reconstruction results. An MLV map of an image can be represented as follows:
\begin{equation}
\label{mlv1}
\begin{aligned}
  \Phi(\hat{\mathbf{y}})=
\begin{bmatrix}

\phi(\hat{y}_{1,1})& \cdots& \phi(\hat{y}_{1,N}) \\
\vdots & \ddots & \vdots\\
\phi(\hat{y}_{M,1})& \cdots & \phi(\hat{y}_{M,N})
\end{bmatrix},\\ 
\end{aligned}
\end{equation}
where $\phi(\hat{y}_{i,j})$ can be calculated as below,
\begin{equation}
\label{mlv2}
    \phi(\hat{y}_{i,j}) = \max|\hat{y}_{i,j} - \hat{y}_{x,y}|,
\end{equation}
where $x= i-1,i,i+1$ and $y = j-1,j,j+1$. From Eq. \ref{mlv1} and Eq. \ref{mlv2}, the MLV of a pixel $\hat{y}_{i,j}$ is the maximum variation between the intensity of a pixel $\hat{y}_{i,j}$ and its 8-neighbor pixels. 

\textbf{Discussion.} As shown in Figure \hyperlink{mlv_figre}{3}, the rationality of MLV-based sharpness-aware distribution alignment comes from several aspects: 1) The sharpness as second-order information of images can capture the high-frequency content of an image, while the original reconstruction images may contain a mixture of content and style information. For example, by observing Figure \hyperlink{mlv_figre}{3g} and \hyperlink{mlv_figre}{3i}, there is still an obvious distribution discrepancy between NDCT images on the source and target domains, even if the content information of LDCT images on the target domain has been reconstructed well. (2) By using sharpness, the interference of content information is reduced  and the focus is shifted to the high-frequency components that are more indicative of the style information. For example, we can observe from Figure \hyperlink{mlv_figre}{3j} that the histogram distribution is highly consistent if two images have similar styles with different content information. This can make adversarial learning more effective by reducing the potential confusion between style and content information, leading to better alignment of the conditional distributions. 3) Theoretically, an MLV map can be regarded as a projection on a joint Gaussian and hyper-laplacian distribution for an image \cite{bahrami2014fast}, which means that the content information of an image can be suppressed using the variation of the content information (see Figure \hyperlink{mlv_figre}{3h}). 

Then, a sharpness-aware distribution alignment (SDA) is proposed in an adversarial learning manner. As shown in Figure \hyperlink{framework}{2b}, a parameterized discriminator $R$ can be utilized to distinguish 
whether the sharpness level of reconstruction results from the target domain is similar to that of NDCT images from the source domain.
The significant distribution discrepancy of MLV maps between the two domains will induce a high discriminator loss, which would push the reconstruction model $F$ to reduce such a sharpness gap, leading to more high-quality reconstruction results for the target domain. Here, we adopt the LSGAN\cite{mao2017least}-based adversarial learning process due to its more stable convergence, which can be formulated as follows,
\begin{eqnarray}
\label{adver}
\label{adversarial}
    \min_{\theta_{R}} \mathcal{L}(R) &=& \mathbb{E}_{\mathbf{y}^{S}\sim p(\mathbf{y}^{S})}[R[\Phi(\mathbf{y}^{S})] - 1] \nonumber\\ &+&\mathbb{E}_{\mathbf{x}^{T}\sim p(\mathbf{x}^{T})}[R[\Phi(G(E(\mathbf{x}^{T})))]- 0)], \nonumber \\
    \min_{\mathbf{w}_{E}, \mathbf{w}_{G}}   \mathcal{L}(E,G) &=& \mathbb{E}_{\mathbf{x}^{T}\sim p(\mathbf{x}^{T})}[R[\Phi(G(E(\mathbf{x}^{T})))] - 1 ]. \nonumber \\
\end{eqnarray}
Eq. \ref{adver} can encourage the model to reconstruct high-quality CT images for target domains with a similar level of sharpness as the NDCT images from the source domain.
\begin{algorithm}[!t]  
  \caption{UDA for LDCT Image Reconstruction}  
  \label{algorithm}  
  \begin{algorithmic}[1]  
    \Require \\
      Paired LDCT and NDCT images on source domain $D_{S} = \{(\mathbf{x}_{1}^{S},\mathbf{y}_{1}^{S}),\cdots, (\mathbf{x}_{N}^{S},\mathbf{y}_{N}^{S}))\}$; Testing LDCT images on target domain $D_{T} = \{\mathbf{x}_{1}^{T},\cdots, \mathbf{x}_{L}^T\}$. \\ Initial Bayesian encoder $E$, Bayesian decoder $G$, and the discriminator $R$, and self-reconstruction network $SR$.
    \Ensure  
      Learned Bayesian encoder $E$ and decoder $G$
     \State Sample a mini-batch $\mathbf{X}^{S}\leftarrow\{\mathbf{x}_{i}^{S}\}_{i=1}^{B}$ and corresponding $\mathbf{Y}^{S}\leftarrow\{\mathbf{y}_{i}^{S}\}_{i=1}^{B}$, and $\mathbf{X}^{T}\leftarrow\{\mathbf{x}_{i}^{T}\}_{i=1}^{B}$ 
     
     \State Compute the $\mathcal{L}_{SL}$ according to Eq. \ref{elbo} based on $\mathbf{X}^{S}$, $\mathbf{Y}^{S}$, and Bayesian reconstruction network $E\circ G$.
      \State Calculate the $\mathcal{L}_{BUA}$ according to Eqs. \ref{bnua} and \ref{self} based on $\mathbf{X}^{S}$, $\mathbf{X}^{T}$, the Bayesian encoder $E$, the Bayesian decoder $G$, and self-reconstruction network $SR$.
      \State Calculate the $\mathcal{L}_{SDA}$ according to Eq. \ref{adversarial} based on $\mathbf{X}^{T}$, $\mathbf{Y}^{S}$, Bayesian reconstruction network $E\circ G$,  the discriminator $R$, and sharpness descriptor $\Phi$ in Eq. \ref{mlv1}.
      \State Update the $E, G, R,$ and $SR$' parameters using the total of $\mathcal{}{L}_{SL},\mathcal{L}_{BUA}$ and $\mathcal{L}_{SDA}$ in Eq. \ref{all}.
      \State Output well-learned Bayesian encoder $E$ and decoder $G$ for testing LDCT image reconstruction.
  \end{algorithmic}  
  \label{algo}
\end{algorithm}

\subsection{Model Implementation}
\label{Overall Framework}
Our proposed UDA framework for LDCT reconstruction problem 
consists of four modules, including a probabilistic encoder, a probabilistic decoder, a self-reconstruction network, and a discriminator. For the probabilistic encoder-decoder network, we follow the structure of the popular CT reconstruction backbone, \textit{i.e.}, a widely-adopted CPCE model \cite{shan20183}.
To balance the computational consumption with the sufficient probabilistic property, we follow \cite{anonymous2023domain} to only replace the last layer of the encoder and the decoder with its Bayesian neural network version. Specifically, the last layer of the encoder is a Bayesian-based CNN layer with 32 channels and $3 \times 3$ kernel size. The last layer of the decoder is a Bayesian-based deconvolutional layer with 1 channel and $3 \times 3$ kernel size. Note that the number of weight parameters of these two layers is twice that of the ordinary CNN layer. The structure of the discriminator in the sharpness-aware distribution alignment module is the same as that of LSGAN \cite{zhu2017unpaired}. Specifically,  the discriminator consists of 4 sequential convolutional learning units, where each unit has 5$\times$5 convolutional layer, a batch normalization layer, and a Leaky ReLU activation unit. The number of convolutional channels is 64, 128, 256, and 512, respectively. Note that the number of strides is 2 for all convolutional layers. Finally, a fully connected layer is used to output a single scaler, \textit{i.e.,} the number of neurons
in the fully connected output layer is set to 1. The overall objective of our proposed framework in the probabilistic reconstruction network can be represented as follows: 
\begin{eqnarray}
\mathcal{L}&=&  \beta_{1} \mathcal{L}_{SL}+ \beta_{2}\mathcal{L}_{BUA} + \beta_{3}\mathcal{L}_{SDA},\label{all}
\nonumber \\
\end{eqnarray}
where the first term denotes the source domain-related loss, which can be represented as follows:
\begin{eqnarray}
    \mathcal{L}_{SL} &=& \sum_{i}[\|\mathbf{y}_{i}^{S} - \hat{\mathbf{y}}_{i}^{S}\|_{1}+PL(\mathbf{y}_{i}^{S},\hat{\mathbf{y}}_{i}^{S})] \nonumber\\&+& \operatorname{KL}[q_{\theta}(Q_{\phi})  \| p(Q_{\phi})]  + \operatorname{KL}[q_{\theta}(C_{\omega})\|p(C_{\omega})] \label{elbo}
\end{eqnarray}
where $Q_{\phi}$ and $C_{\omega}$ denote the Bayesian parameters of the last layer in the encoder $E$ and the decoder $G$. Eq. \ref{elbo} can be regarded as an ELBO loss, which is guided from both image space (\textit{i.e.}, the mean absolute error) and feature space (\textit{i.e.}, the perceptual loss \cite{johnson2016perceptual} $PL(\cdot,\cdot)$). 
Note that the variance of log-likelihood is set to 1 for simplification, computed by the ground-truth $\mathbf{y}_{i}$ and its estimation $\hat{\mathbf{y}}_{i}$. The third and fourth terms in Eq. \ref{elbo} aim to learn a variational distribution $q_{\theta}(\cdot)$ to approximate the Bayesian posterior distribution on the weights while minimizing the KL divergence with its prior distribution $p(\cdot)$. $\beta_{1}$, $\beta_{2}$ and $\beta_{3}$ in Eq. \ref{all} control influences of the source domain loss, the BUA module, and SDA module, respectively. The overall algorithm of our proposed method can be found in Algo. \ref{algo}. Our code is available at \href{https://github.com/tonyckc/UDA-BUA}{https://github.com/tonyckc/UDA-BUA}.

%\begin{table}[!t]
%\caption{The details of used datasets for the validation of the robust cross-domain CT %reconstruction task.}
%\setlength{\extrarowheight}{3pt}
%\begin{adjustbox}{width=0.5\textwidth}
%\begin{tabular}{cc|ccc}
%\hline
%                        & \textbf{Source Domain} & \multicolumn{3}{c}{\textbf{Target Domain}}        %        \\ \hline
%\textbf{Dataset}        & \TEXTTT{AAPM-16}                & Ultra-dose \TEXTTT{AAPM-16} & High-dose AAPM-16 & ISICDM-20        \\
%\textbf{Type}           & Simulated LDCT         & Simulated LDCT     & Simulated LDCT    & Clinical         \\
%\textbf{Radiation Dose} & $\sim$25\%             & $\sim$5\%          & $\sim$50\%        & 120kVp 42$\sim$85mAs  \\
%\textbf{Resolution}     & $512 \times 512$       & $512 \times 512$   & $512 \times 512$  & $512 \times 512$ \\
%\textbf{Pair Number}    & 4800                   & 1136               & 1136              & 1124             \\ \hline
%\end{tabular}
%\end{adjustbox}
%\label{dataset}
%\end{table}
\begin{table}[]
\caption{The details of used datasets.}
\centering
\begin{adjustbox}{max width=\columnwidth}
\begin{tabular}{ccccc}
\hline
\textbf{Dataset}         & \textbf{AAPM-16} &   \textbf{AAPM-A} & \textbf{AAPM-B} & \textbf{ISICDM-20} \\ \hline \hline
 Domain Type  & Source            & Target & Target  & Target \\
Dose Level          & $\sim$25\% &   $\sim$5\%   &  $\sim$50\%   & $\leq$25\%\\
Slice thickness & $\leq$1mm              &  $\leq$1mm   &     $\leq$1mm  & 2mm\\
Resolution &   $512 \times 512$           & $512 \times 512$     & $512 \times 512$ & $512 \times 512$\\
Device           & Siemens             &  Siemens   &  Siemens & UIH\\
\#Total Slice          & 4800             &  1136  &   1136   & 3275 \\ \hline   
\end{tabular}
\end{adjustbox}
\label{dataset}
\end{table}
\section{Experiments and Analyses}
%In this section, extensive experiments are introduced 

\subsection{Datasets and Training Protocol}
\textit{ 1) Mayo 2016 Dataset (Source Domain):} ``2016 NIH-AAPM-Mayo Clinic Low-dose CT Grand Challenge"  dataset (namely \texttt{AAPM-16} for short) is the widely-adopted benchmark dataset for LDCT reconstruction methods. \texttt{AAPM-16} dataset provides 5936 pairs of LDCT/NDCT images with 1mm thickness from 10 patients, where the LDCT images are simulated by inserting the Poisson noise into the projection data (before the image reconstruction) and the noise level of LDCT images corresponds to \textbf{25\%} of the NDCT images. Most CT images are acquired and reconstructed by \textit{Siemens} CT scanners from German. In light of the publicity and accessibility, we adopt the \texttt{AAPM-16} dataset as the source domain and transfer the knowledge of this dataset into the related yet different target domains. Here, we randomly select 8 patients as the paired source domain, \textit{i.e.} 4800 pairs of LDCT/NDCT images.

\textit{2) AAPM-A Dataset (Target domain):} In clinical settings, LDCT images collected from target domains may encounter greater levels of noise interference compared to the source domain. This is due to various factors such as decreased radiation dose, higher slice thickness, larger pixel size, or the use of an outdated CT scanner, which are common sources of domain gaps. For the sake of simplification, we use the remaining 2 patients (1136 pairs of LDCT/NDCT images) from the \texttt{AAPM-16} dataset to directly simulate such heavier noise scenarios by inserting the widely-recognized mixed Poisson-Gaussian noise into the projection data of NDCT, as follows:
%one of the most representative domain gaps is the increase of noise level on the target domain, as the type of noise distribution is usually regarded as complex Gaussian-Poisson noise. Diverse variations will cause higher noise levels in the image domain,  such as the decreased radiation dose, a higher slice thickness, a larger pixel size, and an outdated CT scanner. For the sake of simplification, we simulate an ultra-low-dose imaging scenario to validate the effectiveness of DA methods in increased noise level. insert the widely-recognized mixed Poisson-Gaussian noise into projection data of NDCT to simulate ultra-low-dose imaging scenarios with increased noise level, as follows:
\begin{equation}
    p_{ld} = \operatorname{ln}\frac{I_{0}}{\operatorname{Poisson}(I_{0}\exp(-p_{hd}))+\operatorname{Gaussian}(0,\sigma_{e}^{2})}, \label{noise simulation}
\end{equation}
where $p_{ld}$, $p_{hd}$, $\sigma_{e}^{2}$, and $I_{0}$ denote the low-dose project, normal-dose projection, the variance of electronic noise, and the number of incident photons, respectively. We set $\sigma_{e}^{2}$ and $I_{0}$ to 10 and 1.2$\times 10^{5}$. Then, LDCT images are obtained by the filtered back projection (FBP) algorithm. Due to the heavier noise, the mean peak signal-to-noise ratio (PSNR) of reconstruction images decreases to around 24 dB (approximates 5\% dose of the NDCT images),  significantly lower than that of the source domain (about 28 dB). We coin this dataset as \texttt{AAPM-A}. 

\textit{3) AAPM-B Dataset (Target domain):} It is also common that the collected LDCT images from target domains have milder noise interference than the source domain, due to increased radiation dose, a thinner slice thickness, or an advanced CT scanner. We adopt a similar noise model in Eq \ref{noise simulation} to simulate such imaging scenarios, where $\sigma_{e}^{2}$ and $I_{0}$ are set to 10 and 1.2$\times 10^{6}$, respectively. The mean PSNR of reconstruction images decreases to about 29 dB (approximates 5\% dose of the NDCT images), higher than that of the source domain.
This dataset is coined as AAPM-B.

\textit{4) ISICDM-20 Dataset (Target domain):} ``2020 ISICDM Challege for the Quality Optimization for Low-dose CT Image" dataset (namely \texttt{ISIDM-20} for short) refers to the clinical low-dose imaging datasets. Specifically, 10 patients (3275 slices in total) were scanned by low-dose imaging mode, and another 10 patients were scanned by high-dose imaging mode. Thus, provided LDCT/NDCT images are unpaired. The variations between \texttt{AAPM-16} dataset and \texttt{ISICDM-20} dataset are diverse, such as the changes in the CT manufacturer (\textit{United Imaging Healthcare} (UIH) CT scanners from China), reconstruction kernel, and the pixel spacing. By observing the LDCT images between the two domains, the noise level of \texttt{AAPM-16} is roughly lower than that of \texttt{ISICDM-20}.

\begin{table}[!t]
\centering
\caption{ \textbf{25\% dose level $\rightarrow$ 5\% dose level:} The experimental results on the \texttt{AAPM-A} dataset, \textit{i.e.}, the \texttt{AAPM-16}  dataset as the source domain, and the  \texttt{AAPM-A} dataset as the target domain. The average value and standard deviation are reported by running each method with five times. For PSNR and SSIM, the higher, the better. For GMSD and DSS, the lower, the better. The best and second-best performance in each column is colorized by the \textcolor{red}{red} and the  \textcolor{blue}{blue}. }
\setlength{\extrarowheight}{1pt}
\begin{adjustbox}{width=0.5\textwidth}
\begin{tabular}{cccll}
\hline
\textbf{Method} & \textbf{PSNR} $\uparrow$     & \textbf{SSIM} $\uparrow$     & \multicolumn{1}{c}{\textbf{GMSD} $\downarrow$}  & \multicolumn{1}{c}{\textbf{DSS} $\downarrow$}  \\ \hline \hline
%\textbf{FBP}    & 19.4211            & 0.5926            & %\multicolumn{1}{c}{7.5777}       & \multicolumn{1}{c}{0.2249}    %   \\
\textbf{FBP} & 24.7257 &0.7107 &  1.1926&  0.2647 \\ \hline
\textbf{BM3D} & 29.6162 & 0.8126 & 0.8595& 0.1799 \\
\textbf{ONLM} & 28.0737 & 0.7971 & 0.9574 & 0.2085 \\ 
\textbf{Noiser2Noise} & 26.7995  & 0.7434  & 1.0948  & 0.2383  \\
\textbf{CycleGAN} & 30.1217 & \textcolor{blue}{0.8386}  & 0.7693 & 0.1944 \\
\textbf{CCDnet} & 30.2532& 0.8288 & \textcolor{blue}{0.7255} & 0.1776\\
\textbf{UDA} & \textcolor{blue}{30.8746} & 0.8307 & 0.8661 & 0.1662 \\
\textbf{N2N-R} & 28.1526 & 0.8152 & 0.8957 & 0.2142 \\ \hline \hline
%\textbf{MF-BNN} &  20.9819$\pm$3.2014 & 0.6284$\pm$0.0312 & 5.5392$\pm$0.0125  &  0.2230$\pm$0.0002\\
\textbf{Backbone} & 28.5636 &0.7478 & 
0.6387 & \textcolor{blue}{0.1629} \\  
\textbf{Backbone+DA (Ours)}  & \scalebox{1.1}{\textcolor{red}{\textbf{32.6840}}}& \scalebox{1.1}{\textcolor{red}{\textbf{0.8726}}} & \scalebox{1.1}{\textcolor{red}{\textbf{0.5380}}}                & \scalebox{1.1}{\textcolor{red}{\textbf{0.0947}}}               \\ \hline 
\end{tabular}
\end{adjustbox}
\label{aapm5}
\end{table}

\subsection{Implement Details}
The proposed UAD framework for LDCT reconstruction is implemented by Pytorch and BayesianTorch\footnote{https://github.com/IntelLabs/bayesian-torch}. We adopt mean-field  variational inference (MFVI)\cite{graves2011practical} for the analytical approximation of the Bayesian layer, where the parameters are characterized by fully factorized Gaussian distribution endowed by variational parameters $\mu$ and $\sigma$, \textit{i.e.} $q_{\theta}(w) := \mathcal{N}(w|\mu,\sigma).$ By using BayesianTorch, deterministic neural networks can be transformed easily into their Bayesian versions. For the parameter settings of Bayesian weight priors,  empirical Bayes using DNN (MOPED) \cite{krishnan2020specifying} method is used, where the initial perturbation factor $\delta$ for the weight to 0.1. The number of MC sampling during training is set to 10 empirically. The parameters of the overall model are updated by Adam optimizer with the learning rate as $1\times 10^{-4}$. The number of the mini-batch is set to 8 with a patch size of 256$\times$256. All datasets have been preprocessed to clip between [-1000,400] Hounsfield Unit (HU) and normalized to [0,1]. The epoch is set to 50.  During the training phase in each cross-domain task,  \texttt{AAPM-16} dataset is the source domain (using paired LDCT/NDCT images) and \texttt{AAPM-A}/\texttt{AAPM-B}/\texttt{ISICDM-20} dataset is respectively as the target domain (using LDCT images only). The important hyperparameters (including $\beta_{1}$, $\beta_{2}$, and $\beta_{3}$) and model selection are conducted by test-domain validation by following \cite{zhang2023adanpc}. Specifically, for \texttt{AAPM-A}, \texttt{AAPM-B}, and \texttt{ISICDM-20} datasets, $\beta_{1}$ is set to 1, 0.01, and 0.1. $\beta_{2}$ is set to 10, 1, and 1. $\beta_{3}$ is set to 100, 1, and 0.1.

%\begin{figure}[!t]
%    \centering
%  \includegraphics[width=0.5\textwidth]{figs/sample.pdf}
%  \caption{Example samples of the used abdomen and head datasets. The noise map of each LDCT image (in the third %column) is obtained by the subtraction between the LDCT image and corresponding NDCT image. The display windows for %the head scan and the abdomen scan are [-10,80] and [-160,240], respectively.}
%  \label{different region}
%\end{figure}

\subsection{Baseline Methods and Evaluation Metrics}
\textit{1) Baseline Methods:} In this paper, we compare our proposed method with some representative CT reconstruction approaches, including Block-matching and 3D filtering (BM3D) method \cite{dabov2006image}, optimized non-local means (ONLM) method for CT reconstruction \cite{kelm2009optimizing}, Noiser2noise \cite{moran2020noisier2noise}, CycleGAN \cite{zhu2017unpaired}, unsupervised domain adaptation method for LDCT denoising (UDA)  \cite{lee2022unsupervised}, local and global information alignment-based cross-domain LDCT denoising (CCDnet) method \cite{huang2022cross}. Note that BM3D and ONLM are traditional model-based LDCT image reconstruction methods. Noiser2noise only refers to the LDCT images and corresponding noisier versions on the target domain to conduct unsupervised image denoising. By utilizing the LDCT images on the target domain and the NDCT 
images on the source domain, CycleGAN can learn a transferable style from target domains to source domains \cite{zhu2017unpaired}. Note that 
we also introduce an MFVI-based Bayesian neural network (namely ``Backbone")\cite{20221111787255} as a baseline model, which is only trained by paired source domain data. The network structure of this model is equal to that of our proposed method, except for the lack of domain adaptation strategies.  The hyperparameters of all baseline methods are tuned in a wide range by test-domain validation.

\textit{2) Evaluation Metrics:}  During the testing phase,
we compute the quantitative performance of reconstruction images by corresponding NDCT images. For \texttt{ISICDM-20} dataset, we follow recent work \cite{li2023low} to compare the average CT value of similar regions of interest (ROI) in the reconstructed LDCT images and referenced NDCT images on the target domain. For \texttt{AAPM-A} and \texttt{AAPM-B} datasets, we introduce two kinds of evaluation approaches, including image-based evaluation metrics and perception-based evaluation metrics. The former 
includes the peak signal-to-noise ratio (PSNR) and the structural similarity index measure (SSIM) \cite{wang2004image}, where PSNR can measure the ratio between the maximum possible power of a signal and the power of the noise, and SSIM can take into account the differences of luminance, contrast, and structural information between two images.
The latter includes the gradient magnitude similarity deviation (GMSD) \cite{xue2013gradient} and discrete cosine transform-based sub-bands similarity index (DSS) \cite{zhang2012sr}, which respectively derive from the perspectives of image gradients and structure information, achieving the approximation of the human visual system. For \texttt{ISICDM-20}, the absolute error (AR) between the average CT value of reconstructed images and referenced NDCT images is computed. Note that all quantitative results are calculated   based on a CT window
of [-1000, 400]HU.

\begin{figure*}[!t]
    \centering
  \includegraphics[width=0.99\textwidth]{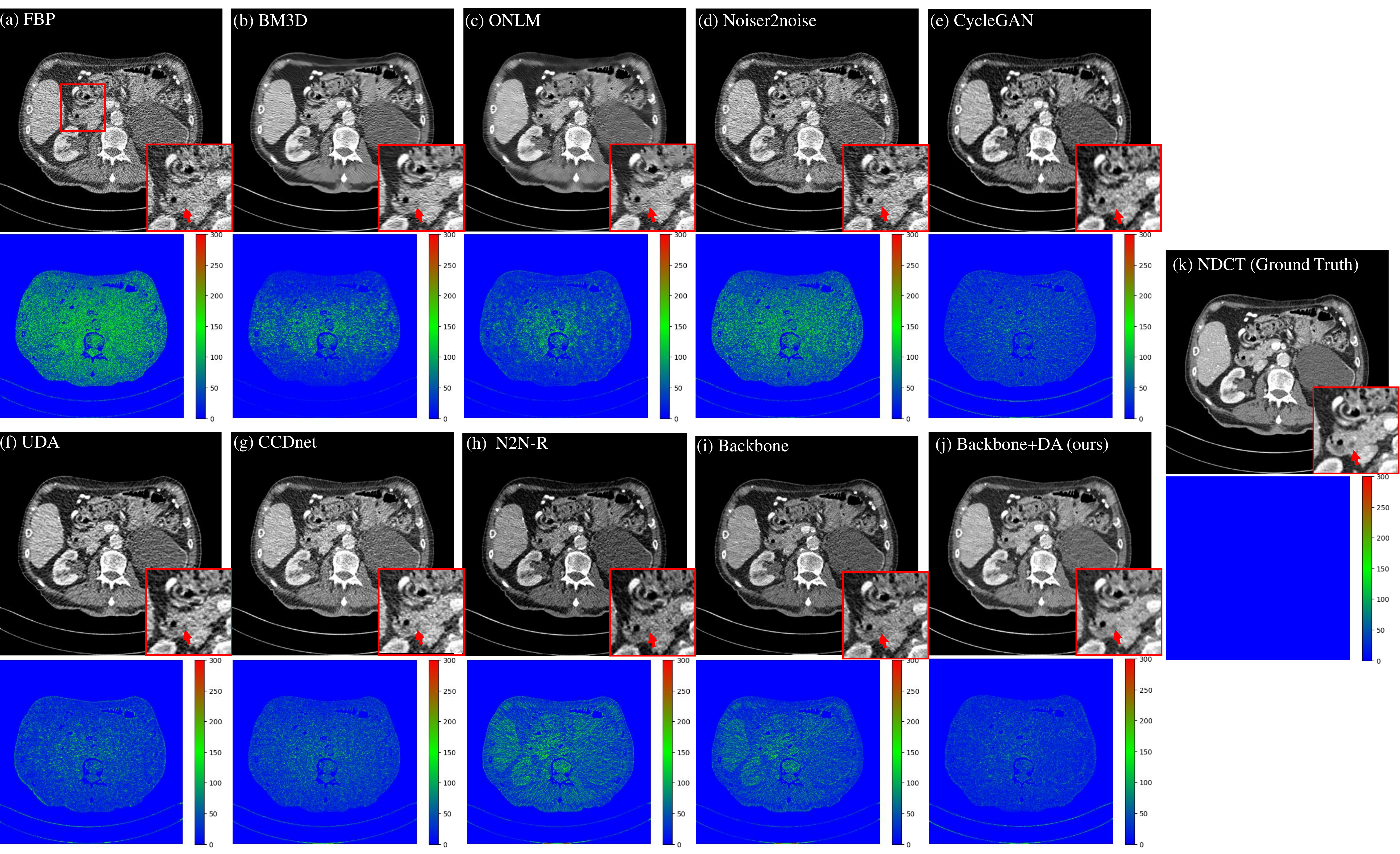}
  \caption{Qualitative results of an example CT image from \texttt{AAPM-A} dataset. The red ROI is zoomed in for visual comparison, and the red arrow points to one lesion. The absolute differences between the reconstructed results and corresponding NDCT images NDCT are shown in the second and last row. The display window is [-160,240] HU.}
  \label{AAPM5_figure}
\end{figure*}
\begin{figure*}[!t]
    \centering
  \includegraphics[width=0.99\textwidth]{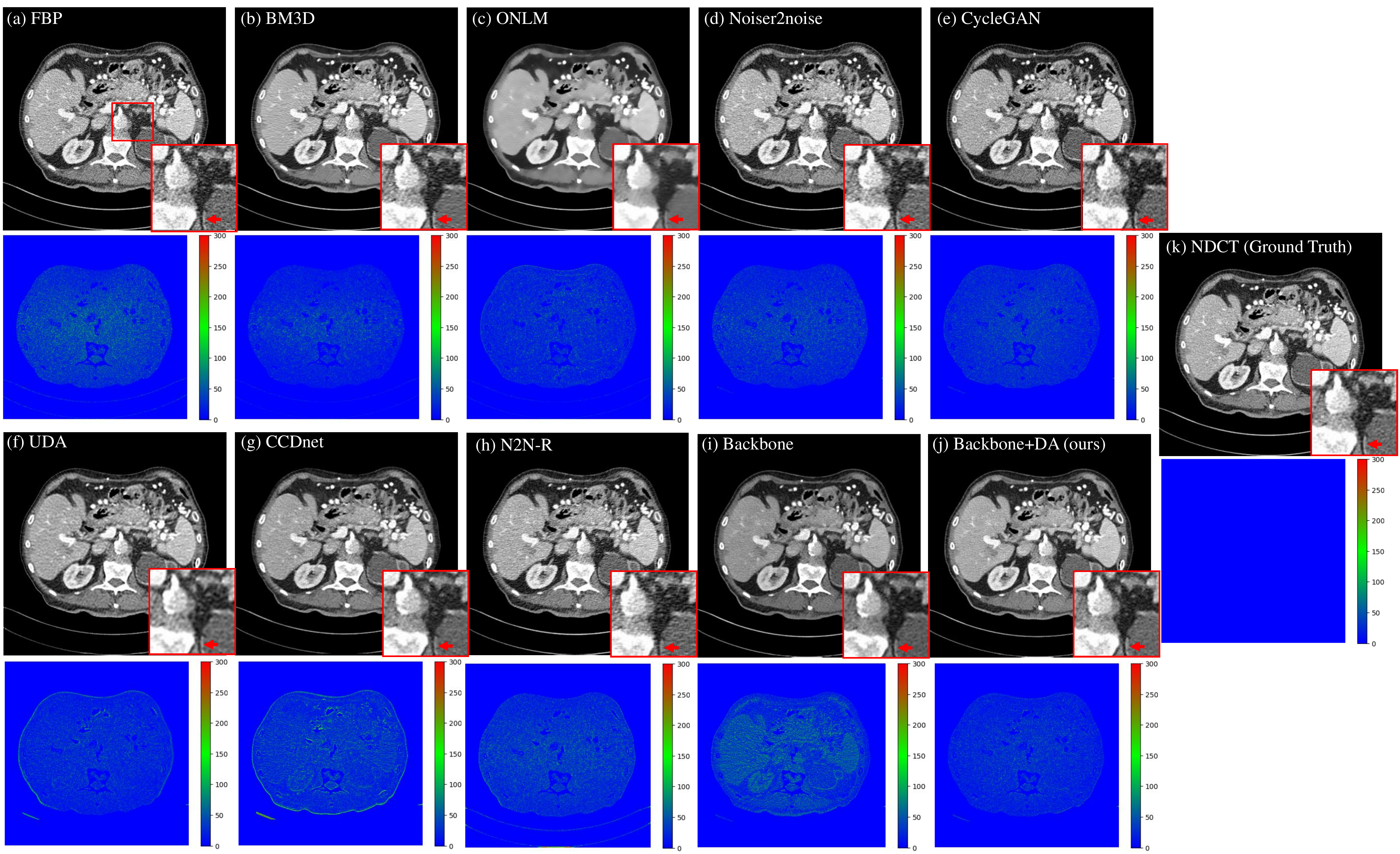}
  \caption{Qualitative results of an example CT image from \texttt{AAPM-B} dataset. The red ROI is zoomed in for visual comparison, and the red arrow points to one lesion. The absolute differences between the reconstructed results and corresponding NDCT images NDCT are shown in the second and last row. The display window is [-160,240] HU.}
  \label{aapm50_figure}
\end{figure*}
 
 \begin{table}[!t]
\centering
\caption{\textbf{25\% dose level $\rightarrow$ 50\% dose level:} The experimental results on the \texttt{AAPM-B} dataset, \textit{i.e.}, the \texttt{AAPM-16}  dataset as the source domain, and the  \texttt{AAPM-B} dataset as the target domain. The average value and standard deviation are reported by running each method five times. For PSNR and SSIM, the higher, the better. For GMSD and DSS, the lower, the better. The best and second-best performance in each column is colorized by the \textcolor{red}{red} and the  \textcolor{blue}{blue}. }
\setlength{\extrarowheight}{1pt}
\begin{adjustbox}{width=0.5\textwidth}
\begin{tabular}{cccll}
\hline
\textbf{Method} & \textbf{PSNR} $\uparrow$     & \textbf{SSIM} $\uparrow$     & \multicolumn{1}{c}{\textbf{GMSD} $\downarrow$}  & \multicolumn{1}{c}{\textbf{DSS} $\downarrow$}  \\ \hline \hline
%\textbf{FBP}    & 19.4211            & 0.5926            & %\multicolumn{1}{c}{7.5777}       & \multicolumn{1}{c}{0.2249}    %   \\
\textbf{FBP} & 29.8736 &0.8218 & 0.7074 & 
0.0459 \\ \hline
\textbf{BM3D} & \textcolor{blue}{33.4311} & 0.9283 & \textcolor{blue}{0.3821}& 0.0705 \\
\textbf{ONLM} & 33.2433 & 0.8979 & 0.8226 & 0.1807\\
\textbf{Noiser2Noise} & 32.2477  & 0.8934  & 0.4461  & \textcolor{blue}{0.0617}  \\
\textbf{CycleGAN} & 32.2655 & 0.9116  & 0.3957 & 0.0705 \\
\textbf{CCDnet} & 29.3600& 0.8791& 0.9078 & 0.1339\\
\textbf{UDA} & 31.9557 & \textcolor{blue}{0.9353}& 0.7753 & 0.1718 \\ 
\textbf{N2N-R} & 32.0123 & 0.9032& 0.5268 & 0.0789 \\ \hline\hline
\textbf{Backbone} &  30.2216 & 0.8422 & 0.4273  &  0.1417\\
\textbf{Backbone+DA (ours)}   & \scalebox{1.1}{\textcolor{red}{\textbf{35.1125}}} & \scalebox{1.1}{\textcolor{red}{\textbf{0.9455}}} & \scalebox{1.1}{\textcolor{red}{\textbf{0.3022}}}                & \scalebox{1.1}{\textcolor{red}{\textbf{0.0501}}}               \\ \hline
\end{tabular}
\end{adjustbox}
\label{aapm50}
\end{table}

\subsection{Experimental Results}
\subsubsection{Performance comparison on AAPM-A dataset}
 The quantitative performance of reconstructed results on \texttt{AAPM-A} dataset (as the target domain) can be found in Table \ref{aapm5}. Some observations can be noticed as follows: First, traditional model-based LDCT image reconstruction methods, such as BM3D and ONLM, perform well in terms of PSNR and SSIM, but their perception-based performance in terms of GMSD and DSS is not as good as some other methods. Second, CycleGAN, UDA, and CCDnet methods use different strategies to improve the reconstruction performance of LDCT images in terms of some metrics, their performance is not as good as our proposed method.
Finally, our proposed method, which uses domain adaptation strategies, outperforms all other methods in all four metrics (especially for perception-based scores, \textit{e.g.}, DSS) with a clear margin, indicating that it is highly effective in improving the quality of LDCT images in the target domain.

The visual comparison among different baseline approaches is shown in Figure \ref{AAPM5_figure}. First, we can observe from the zoomed-in view that our proposed method not only can suppress the noise well (please see the absolute difference image) but also can maintain the CT value better than other methods with a more clear view (please see the red arrow). Second, although BM3D and ONLM remove the noise well, they significantly lose texture information compared with our proposed method. This also coincides with the quantitative results in Table \ref{aapm5}, \textit{i.e.}, low perception-based scores with high PSNR scores. Third, compared with other UDA methods (\textit{e.g.,} CCDnet), our proposed method has natural textures closer to NDCT images.

\subsubsection{Performance comparison on AAPM-B dataset}
The quantitative results of different baseline methods on \texttt{AAPM-B} dataset are reported in Table \ref{aapm50}. As we can see, all methods (except for CCDnet) achieve obvious improvements in terms of PSNR, which reflects the effectiveness of these methods in such milder noise interference. Note that our proposed method also has the best performance on all four metrics. Meanwhile, our proposed method outperforms the ``Baseline" by a large margin (from \textbf{0.1417} to \textbf{0.0501} in terms of DDS score), which demonstrates the effectiveness of our proposed DA strategies.

The visual comparison among baseline methods is illustrated in Figure \ref{aapm50_figure}. One has the following observations: First, all baseline methods suppress the noise (please see the absolute difference image) and reconstruct more high-quality results. Second, it seems that the ``Baseline" method significantly loses much contrast information compared with our proposed method, which may be reasonable as the target domain has milder noise interference than the source domain, leading to excessive noise removal on the target domain. Third, we can observe the significant loss of edge information for UDA and CCDnet methods through absolute difference images. Instead, our proposed method imposes a joint discrepancy minimization for domain gaps, resulting in a balanced reconstruction performance. 
\begin{figure*}[!t]
    \centering
  \includegraphics[width=0.99\textwidth]{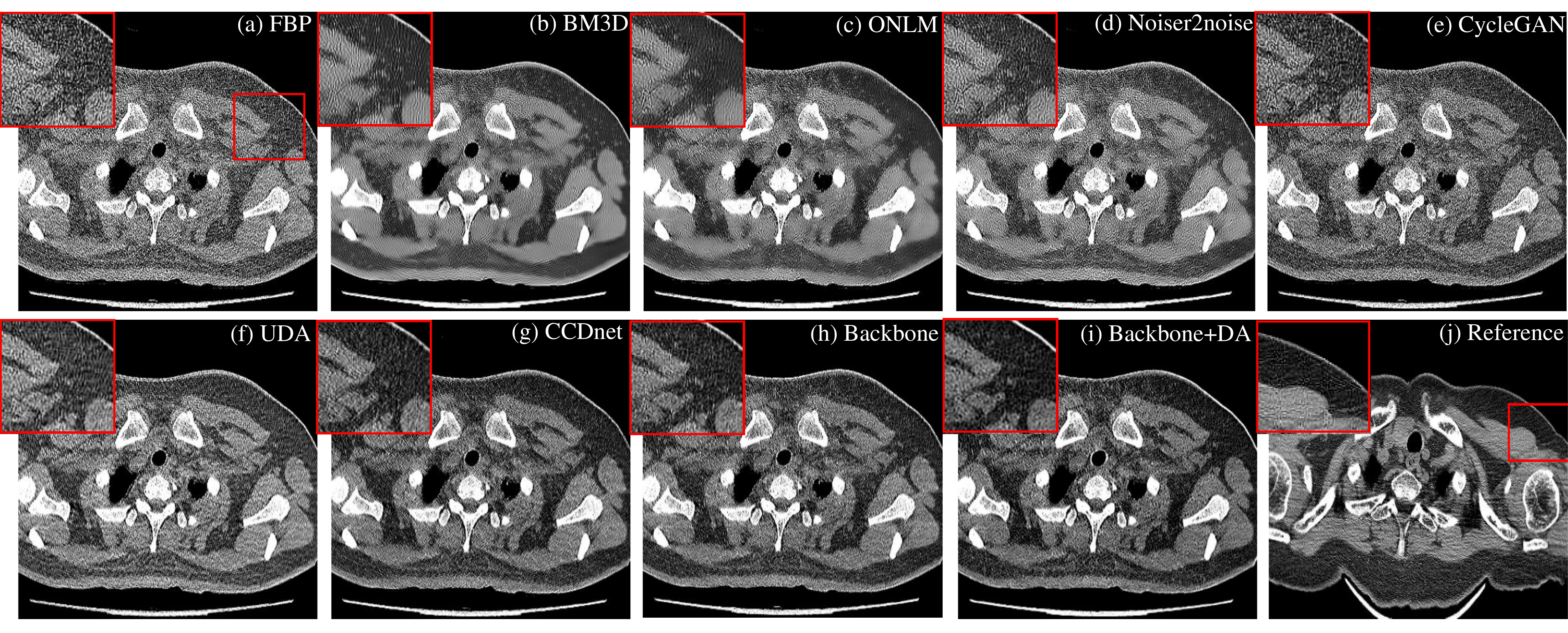}
  \caption{Qualitative results of an example CT from \texttt{ISICDM-20} dataset. The red ROI, namely \textbf{ROI-1}, is zoomed in for visual comparison. The red arrow points to one lesion. The reference image (j) is selected by consideration of the slice similarity from an unpaired NDCT patient in \texttt{ISICDM-20} dataset. Based on the reference image, a similar ROI view is shown. The display window is [-160,240] HU.}
  \label{real1}
\end{figure*}

\begin{figure*}[!t]
    \centering
  \includegraphics[width=0.98\textwidth]{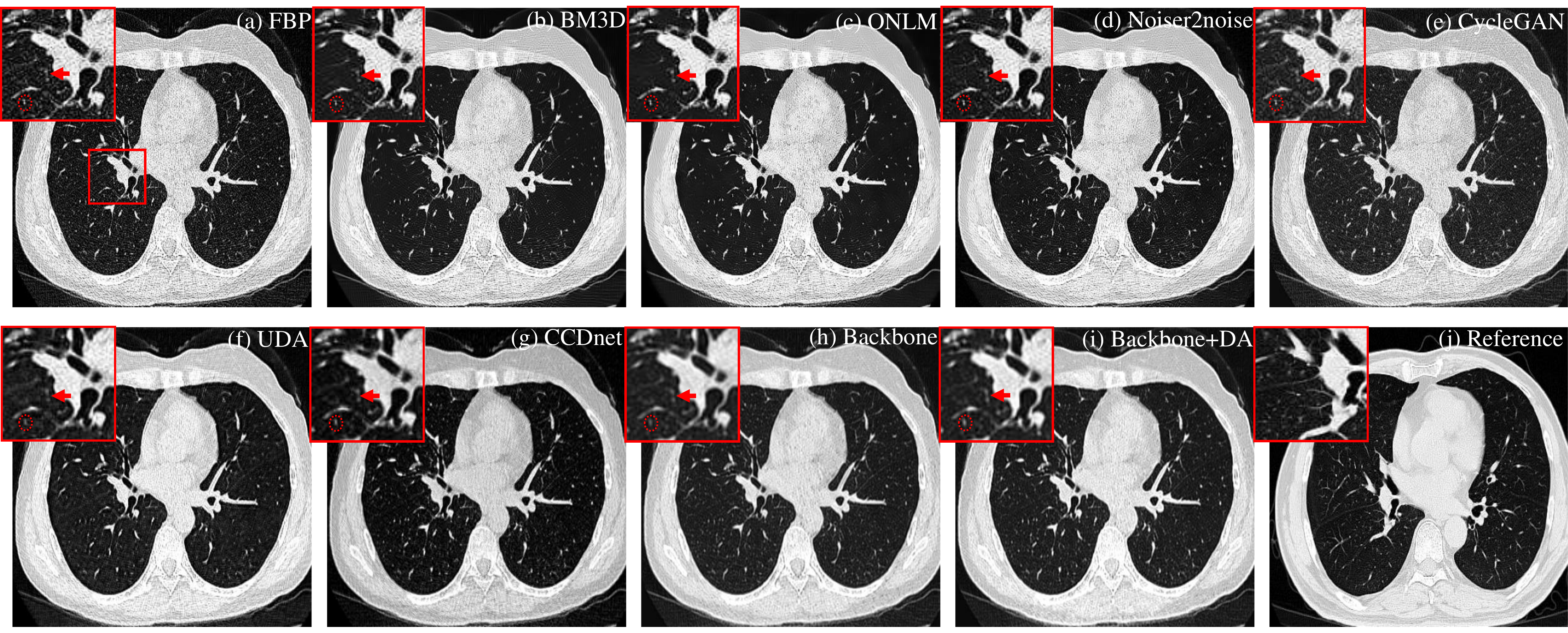}
  \caption{Qualitative results of an example CT from \texttt{ISICDM-20} dataset. The red ROI, namely \textbf{ROI-2}, is zoomed in for visual comparison. The red arrow and circle point to two tiny structures. The reference image (j) is selected by consideration of the slice similarity from an unpaired NDCT patient in \texttt{ISICDM-20} dataset. Based on the reference image, a similar ROI view is shown. The display window is [-1000,150] HU.}
  \label{real2}
\end{figure*}
\begin{table}[!t]
\centering
\caption{Experimental results on the ISIDCM-20 dataset. ``Mean" denotes the average CT value in the selected ROIs. ``AR" denotes the absolute error between the average CT value of each method and that of referenced NDCT in the selected ROIs. The best and second-best performance in each column is colorized by the \textcolor{red}{red} and the  \textcolor{blue}{blue}. }
\setlength{\extrarowheight}{1pt}
\begin{adjustbox}{width=0.4\textwidth}
\begin{tabular}{c|ccll}
\hline
\centering
\multirow{2}{*}{\textbf{Method}} & \multicolumn{2}{c|}{\textbf{ROI-1}}                                    & \multicolumn{2}{c}{\textbf{ROI-2}}                                    \\ \cline{2-5} 
                                 & \multicolumn{1}{c|}{\textbf{Mean}} & \multicolumn{1}{c|}{\textbf{AR} $\downarrow$} & \multicolumn{1}{c|}{\textbf{Mean}} & \multicolumn{1}{c}{\textbf{AR} $\downarrow$} \\ \hline
\textbf{Reference (NDCT)}         & -368.38                          &    0                                         & -28.82                          &    \multicolumn{1}{c}{0}                                         \\ \hline
\textbf{FBP}                      & -335.15                          & 33.23                                        & \multicolumn{1}{c|}{7.29}                            & 36.11                                        \\
\textbf{BM3D}                     & -334.25                          & 34.13                                        & \multicolumn{1}{c|}{8.43}                            & 37.25                                        \\
\textbf{ONLM}                     & -336.84                          & 31.54                                        & \multicolumn{1}{c|}{9.22}                            & 38.04                                        \\
\textbf{Noiser2Noise}             & -336.44                          & 32.94                                        & \multicolumn{1}{c|}{12.25}                           & 40.07                                        \\
\textbf{CycleGAN}                 & -327.20                          & 41.18                                        & \multicolumn{1}{c|}{6.68}                            & 35.50                                        \\
\textbf{CCDnet}                   & \textcolor{blue}{-342.21}                          & \textcolor{blue}{26.17}                                        & \multicolumn{1}{c|}{\textcolor{blue}{-13.91}}                          & \textcolor{blue}{14.91}                                        \\
\textbf{UDA}                      & -338.43                          & 29.95                                        & \multicolumn{1}{c|}{-4.87}                           & 23.95                                        \\ \hline
\textbf{Backbone}                 & -341.25                          & 27.13                                        & \multicolumn{1}{c|}{-52.03}                         & 23.21                                        \\
\textbf{Backbone+DA(ours)}        & \textcolor{red}{\textbf{-352.11}}                          & \textcolor{red}{\textbf{16.27}}                                        & \multicolumn{1}{c|}{\textcolor{red}{\textbf{-25.15}}}                          & \multicolumn{1}{c}{\textcolor{red}{\textbf{3.67}}}                                         \\ \hline
\end{tabular}
\end{adjustbox}
\label{isicdm-20}
\end{table}

\subsubsection{Performance comparison on clinical dataset} We follow \cite{10081080} to choose the two most representative ROIs for performance comparison on clinical datasets, as shown in Figures \ref{real1} and \ref{real2}. We calculate the average CT Hounsfield units value and corresponding absolute error with reference images (unpaired NDCT images) as used in \cite{10081080}.  As illustrated in Table \ref{isicdm-20}, quantitative results demonstrate that our proposed method has a closer CT value with the reference image. For example, in ROI-2, the average CT value of our proposed method is -25.15, instead, the second-best result achieved by CCDnet is only -13.91. By observing visualized results in Figure \ref{real1}, it seems that most baseline methods (especially for Noiser2noise and CycleGAN) achieve an unfavorable noise suppression for this very challenging LDCT image with complex and serious noise.
Although BM3D and ONLM look good for noise removal, they show an unnatural texture compared with our proposed method. From the zoomed-in ROI-1, compared with other deep learning-based methods, our proposed method not only archives the best performance in terms of noise removal (please see the dark region) but also maintains important lesions well (please see the red arrow). There is a similar observation in Figure \ref{real2}. Specifically, our proposed method has better noise suppression capacity compared with Noise2noise and ClycleGAN, while the tiny structure is maintained (as shown in the red circle) well unlike the UDA, CCDnet, and Backbone models. More importantly, our proposed method achieves a reference image-like texture style without obvious artifacts  existing in BM3D and ONLM.

\subsection{Further analysis of our proposed method}
In this section, we further explore the effectiveness of each component in our proposed framework on cross-domain LDCT reconstruction tasks.
%\textit{\textbf{Deterministic Framework v.s. Probabilistic Framework.}} To 
%validate the effectiveness of the probabilistic framework, we compare the deterministic CPCE
%model with our adopted Bayesian version. The quantitative results on the $A\rightarrow B$ task are reported in Table \ref{ablation}. From the first and the second rows of Table \ref{ablation}, we can observe the probabilistic framework outperforms the deterministic framework, especially for the SSIM score, which is reasonable as the probabilistic property of the weights in Bayesian network can act an implicit regularization for better robustness.
\begin{table}[!t]
 \vspace{-0.3cm}
		\centering

	%\resizebox{0.9\columnwidth}{!}{%
		\caption{Ablation analysis of each component of our proposed method. SDA denotes the sharpness-aware distribution alignment module. BUA denotes the Bayesian uncertainty alignment module. CS and CMC denote the channel-wise statistics and covariance matrix computation in the BUA module, respectively. The best and second-best results in each category are bold and underlined. }
 \vspace{0.15cm}
\renewcommand{\arraystretch}{1.2}
\begin{adjustbox}{width=0.5\textwidth}
\begin{tabular}{cccccccc}
\toprule
 & \multicolumn{1}{c}{\textbf{SDA}} & \multicolumn{2}{c}{\textbf{BUA}} & \multicolumn{4}{c}{\textbf{AAPM16} $\rightarrow$ \textbf{AAPM-B}}\\
% \cmidrule(lr){3-4} \cmidrule(lr){5-6} \cmidrule(lr){7-11}
\cmidrule(lr){1-2} \cmidrule(lr){3-4} \cmidrule(lr){5-8}
% Settings &
ID &
 - &
  CS &
  CMC & PSNR $\uparrow$          & SSIM $\uparrow$         & GMSD $\downarrow$           & DSS $\downarrow$       \\ %\midrule
  \midrule
% Baseline &
(a)      & $\times$     & $\times$     & $\times$  & 30.2216          & 0.8422          & 0.4273          & 0.1417     \\
\hline
% \multirow{3}*{\shortstack{Bayesian invariant \\classifier}} &
(b) & $\checkmark$     & $\times$     & $\times$ &          32.2252 & \underline{0.9262} & 0.5661 & 0.1020  \\ 
 (c)  

  &
  $\times$ &
   $\checkmark$&$\times$
   &      32.1242     &   0.8869        &      0.4852     &     0.0896      
   \\
 (d)  &
  $\times$ &$\times$
   &
  $\checkmark$ &           31.9825 & 0.8752 & 0.5089 & 0.0958 \\ 
  (e)      & $\times$ & $\checkmark$     & $\checkmark$  &        \underline{33.8521}       &         0.9188      &       \underline{0.3969}       &     \underline{0.0632}       \\ 
  
% \multirow{3}*{\shortstack{Bayesian invariant\\feature extractor}} &  
 (h)      &  $\checkmark$ & $\checkmark$     & $\checkmark$ &               \textbf{35.1125}       &         \textbf{0.9455}      &       \textbf{0.3022}        &     \textbf{0.0501}      \\ 
% \multirow{3}*{\shortstack{Synergistic Bayesian \\ invariant learning}} & 

  \bottomrule \label{ablation study}
\end{tabular}
\end{adjustbox}
\end{table}

\subsubsection{Ablation Study}
The ablation study of our proposed method can be found in Table \ref{ablation study}. As we can see, our proposed Bayesian uncertainty alignment module (Table \hyperlink{ablation study}{V(e)}) takes more improvements in terms of PSNR, GMSD, and DSS, compared with the sharpness-aware distribution alignment module (Table \hyperlink{ablation study}{V(b)}). By incorporating these two modules into a joint discrepancy minimization, the ``complete model" (Table \hyperlink{ablation study}{V(h)})achieves better performance. 

We further demonstrate the effectiveness of the covariance matrix, we simply compute the  Euclidean distance between channel-wise statistics, which can be regarded as a first-order difference compared with the covariance matrix-based second-order difference. The results can be observed in Table \hyperlink{ablation study}{V(c)} and Table \hyperlink{ablation study}{V(e)}. As we can see, the direct first-order difference in (c) is obviously behind our proposed covariance matrix-based computation in (e), which may be reasonable as the covariance matrix can consider more high-level information between dimensions. Meanwhile, to demonstrate the effectiveness of the channel-wise statistics, we replace global average pooling-based channel-wise statistics with global maximum pooling-based ones. As we can see, our proposed global average pooling-based statistics (Table \hyperlink{ablation study}{V(e)}) outperforms global maximum pooling-based ones (\hyperlink{ablation study}{V(d)}) with a clear margin.
    
\begin{figure}[!t]
    \centering
  \includegraphics[width=0.4\textwidth]{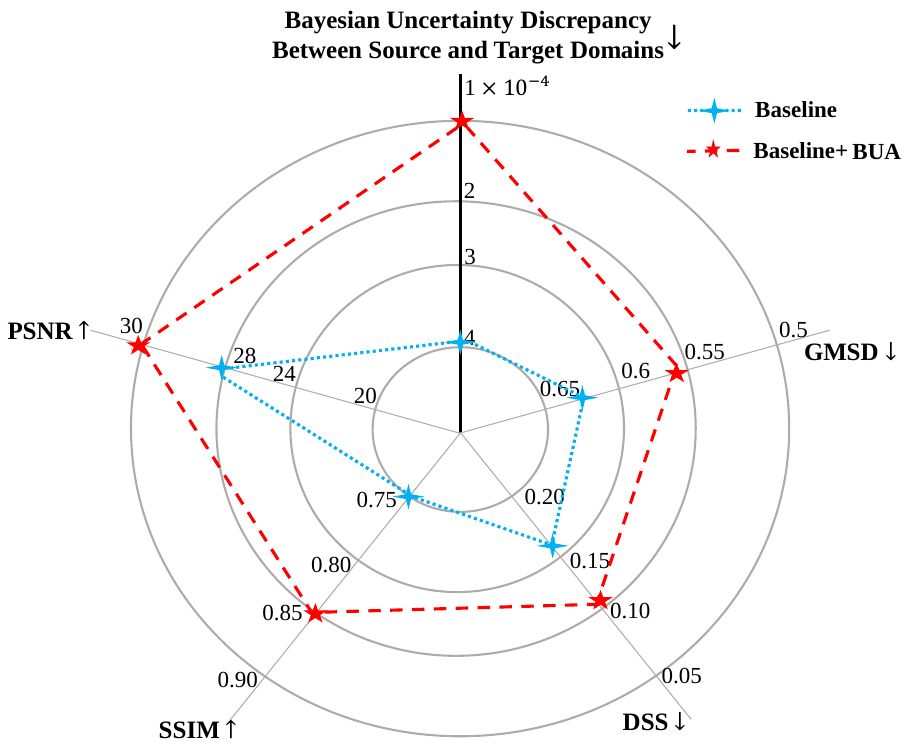}
  \caption{The relationship between Bayesian uncertainty discrepancy (between source and target domains) and quantitative results on \texttt{AAPM-A} dataset. The Bayesian uncertainty discrepancy can be computed using Eq. \ref{bnua}. The average value is reported by running each model 10 times.}
  \label{bnua_discrepancy}
\end{figure}
\subsubsection{Effectiveness of Bayesian Uncertainty Alignment (BUA)} \label{testing_bnua}
BUA module aims to directly reduce the gap in the level of uncertainty between the source and target
domains, making it more likely to render well-reconstructed
results on the target domain. Here, by equipping with our proposed BUA module, we explore \textit{whether the Bayesian uncertainty of the ``unfamiliar" target domain will decrease} and \textit{whether there is a link between the decrease of Bayesian uncertainty and quantitative performance}. To this end, we compute the Bayesian uncertainty discrepancy between source and target domains (as described in Eq. \ref{bnua}) of different models (including the ``Baseline" model and the ``Baseline" + $\mathcal{L}_{BUA}$ model). The quantitative results can be found in Figure \ref{bnua_discrepancy}. As we can see, the baseline model suffers from a huge Bayesian uncertainty discrepancy (see the blue star on the bold black line), which reflects the Bayesian uncertainty discrepancy can implicitly represent the epistemic capacity for those ``unfamiliar" target domain data in the latent space. Instead, by directly reducing the $\mathcal{L}_{BUA}$, the model has a smaller Bayesian uncertainty discrepancy (see the red pentagram on the bold black line), which also contributes to better quantitative results with a clear margin. This is reasonable, as reducing the gap in the level of uncertainty between the source and target domains can contribute to learning a domain-invariant Bayesian model \cite{wen2019bayesian}, leading to the reduction of negative effects of domain shifts between source and target domains.

\subsubsection{First-order information v.s. Second-order Information for Conditional Distribution Alignment} 
\begin{figure}[!t]
    \centering
  \includegraphics[width=0.5\textwidth]{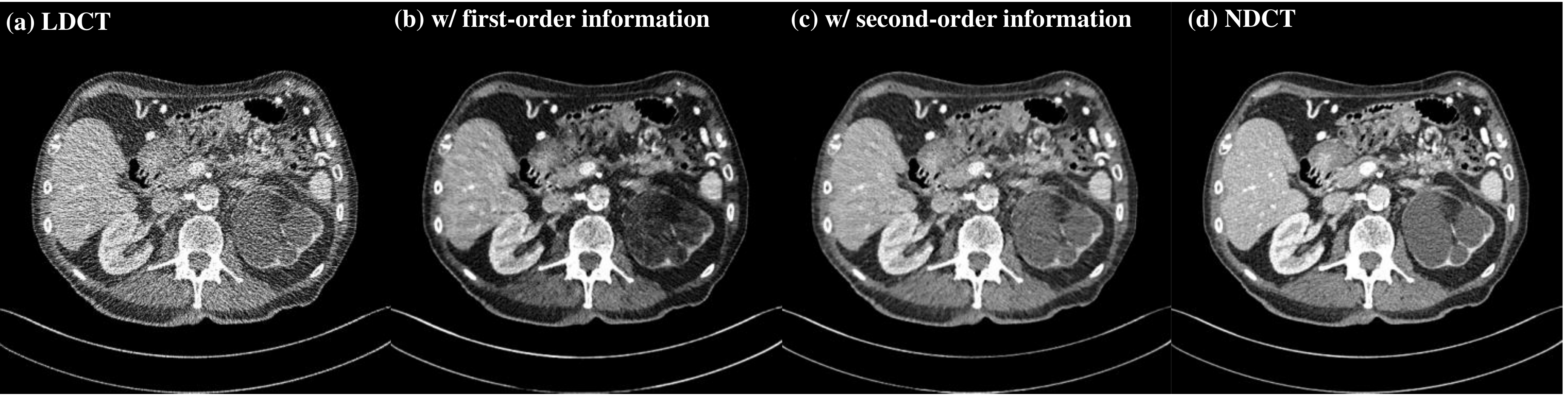}
  \vspace{-0.5cm}
  \caption{Left: Examples from our model with the second-order information-based (\textit{i.e.}, sharpness-aware) distribution alignment (c) and the first-order information-based (original CT images) distribution alignment (d). Right: The learning curve of the discriminator in the adversarial learning process. The display window is [-160,240] HU.   }
  \label{dis_loss}
\end{figure}

We argue that our proposed sharpness-aware distribution alignment (SDA) leverages the sharpness as second-order information of images to conduct conditional distribution alignment between source and target domains, which can effectively alleviate the mismatch issue of content information in the adversarial learning process. To validate this, we remove the MLV operation to evaluate the effectiveness of this first-order information-based adversarial learning. As shown in Figure \ref{dis_loss}, our proposed sharpness-aware distribution alignment can achieve better reconstruction performance by comparing Figures \hyperlink{dis_loss}{9c} and \hyperlink{dis_loss}{9d}. Meanwhile, the learning curve of the discriminator shows that second-order information-based adversarial learning can lead to lower discriminator loss (see Figure \hyperlink{dis_loss}{9}), which may be reasonable as mismatch issue of first-order content information is alleviated by the evaluation of the sharpness.

\subsubsection{Uncertainty Quantification in Cross-domain Scenarios}
\begin{figure}[!t]
    \centering
  \includegraphics[width=.45\textwidth]{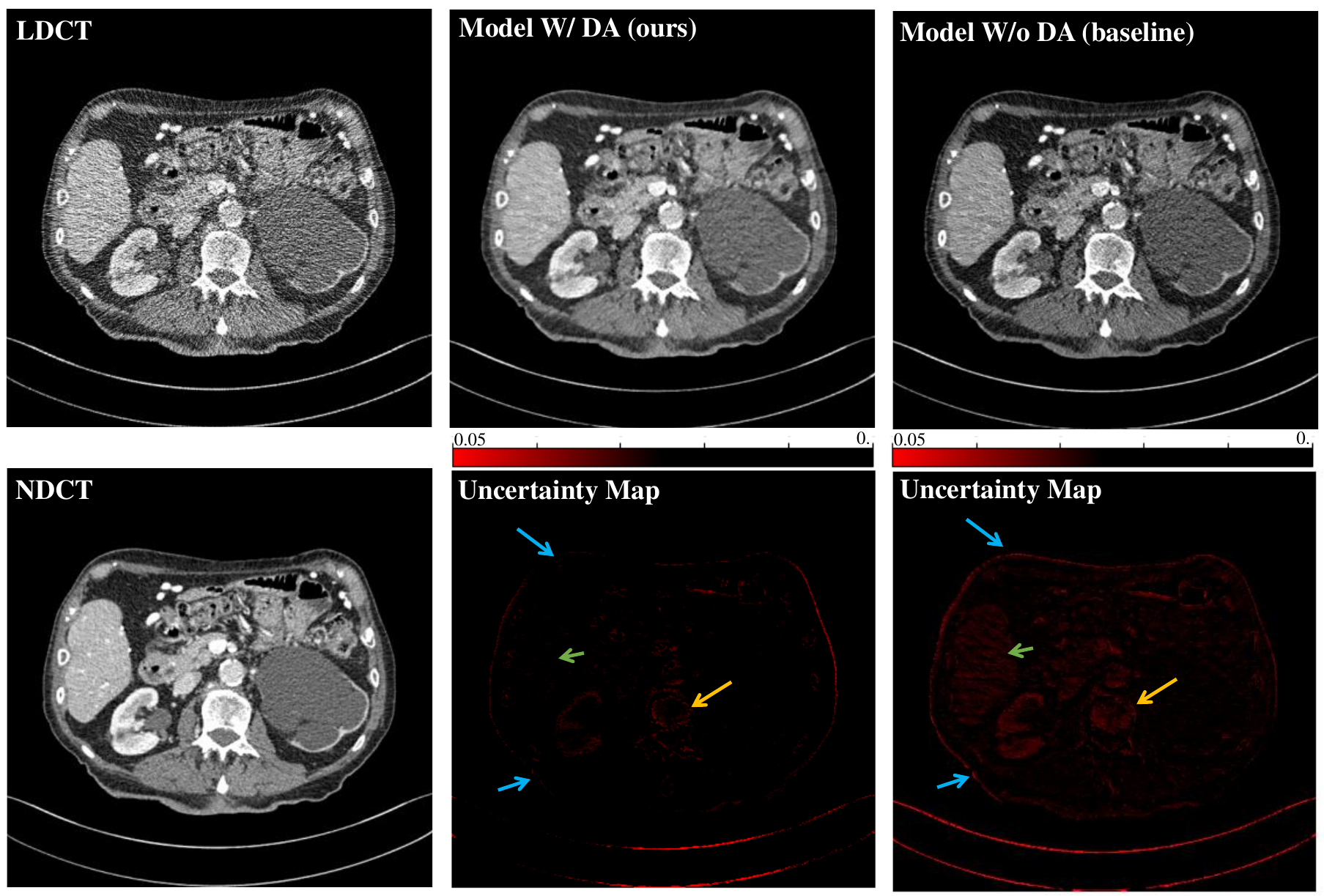}
  \caption{Uncertainty quantification of different models. Note that \texttt{AAPM-A} is the target domain. Example reconstruction results are obtained by our proposed method (with DA strategy) and ``Baseline" model (without DA strategy). If the level of uncertainty is higher, the color in that pixel will be closer to the red.  The display window is [-160,240] HU.   }
  \label{uncertainty}
\end{figure}

Owing to the property of BNN, our proposed method can conduct an effective uncertainty quantification for reconstruction results. Here, as shown in Figure \ref{uncertainty}, we compare the obtained uncertainty map of our proposed method with that of ``Baseline" model to validate the effectiveness of our Bayesian uncertainty alignment for final reconstruction results. Note that the corresponding pixel-wise estimated uncertainty map (on bottom) in the image space is calculated by the variance-based method in Eq. \ref{compute certainty} \cite{cao2020uncertainty,feng2022bayesian}, where the number of MC sampling $T$ is set to 5. We can observe that both our proposed method and the ``Baseline“ method have higher uncertainty on edge regions. However, by our proposed UDA framework, the level of uncertainty of our proposed method decreases significantly (refer to the blue arrow in Figure \ref{uncertainty}). Meanwhile, there is an obvious degradation of the contrast in the  left region (refer to the green arrow in Figure \ref{uncertainty}) by observing the NDCT version. As we can see, the level of uncertainty in this region is higher for the ``Baseline" model, which may coincide with the phenomenon in Figure \ref{motivation}. Finally,  as pointed out by the yellow arrow in Figure \ref{uncertainty}, this region has a high level of uncertainty for the ``Baseline" model,  instead, our proposed method still reports the uncertainty in the edge region. By comparing the LDCT and NDCT images, we can see that the ``Baseline" model introduces additional artifacts. In summary, the uncertainty map not only can reflect the confidence of the model, which may be beneficial to the physicians in the clinic due to the potential correlation with the reconstruction performance but also can validate the effectiveness of our proposed uncertainty-guided method.
\begin{table}[!t]
\centering
\caption{ Ablation study of comparison between CNN-based and transformer-based Bayesian frameworks.  For PSNR and SSIM, the higher, the better. For GMSD and DSS, the lower, the better. }
\setlength{\extrarowheight}{1.2pt}
\begin{adjustbox}{width=0.45\textwidth}
% \usepackage{multirow}
\begin{tabular}{cccccl}
\hline
Type                         & Method      & \textbf{PSNR} $\uparrow$     & \textbf{SSIM} $\uparrow$     & \multicolumn{1}{c}{\textbf{GMSD} $\downarrow$}  & \multicolumn{1}{c}{\textbf{DSS} $\downarrow$} \\ \hline \hline
\multirow{2}{*}{\textbf{CNN}}         & Backbone    &  30.2216 & 0.8422 & 0.4273  &  0.1417     \\
                             & Backbone+DA &  35.1125 & 0.9455 & 0.3022 &\textbf{0.0501}   \\ \hline
\multirow{2}{*}{\textbf{Transformer}} & Backbone    &  31.2341 & 0.8667 & 0.3952  &  0.1287     \\
                             & Backbone+DA &  \textbf{36.2525} & \textbf{0.9598} & \textbf{0.2898} & 0.0689    \\ \hline
\end{tabular}
\end{adjustbox}
\label{transformer-cnn}
\end{table}
\subsubsection{The comparison of different backbone networks} we introduce the \textbf{Hformer}\cite{zhang2023hformer},  a transformer-based LDCT reconstruction model, to play the backbone network of our proposed method. Specifically, we still convert the final layer of the encoder and the decoder of the Hformer to a Bayesian layer. A comparison between the CNN-based and transformer-based frameworks is conducted on AAPM-A dataset. The quantitative results can be found in Table \ref{transformer-cnn}. As we can see, the transformer-based backbone outperforms the CNN-based backbone due to better feature extraction capacity.
\begin{table}[!t]
\centering
\setlength{\extrarowheight}{5pt}
\caption{Average inference time (second(s)) of different methods on overall AAPM-A dataset. The CPU is Intel Xeon Gold 5416S and the GPU is a single NVIDIA-3090. }
\begin{adjustbox}{width=0.5\textwidth}
\begin{tabular}{cccccccccc}
\hline
Methods   & FBP & BM3D& ONLM & Nois2Noise & ClycleGAN & CCDnet & UDA & N2N-R & Ours  \\ \hline \hline
Inference time & -  & 134.56  & 140.52 & 4.42 &  5.51 & 5.22 & 5.18 & 25.65 & 9.85\\ \hline
\end{tabular}
\label{inference time}
\end{adjustbox}
\end{table}
\begin{table}[!t]
\centering
\caption{ Ablation study of different numbers of Bayesian layers in the encoder and decoder. Note that the structure of the encoder and decoder is symmetric. For PSNR and SSIM, the higher, the better. For GMSD and DSS, the lower, the better. }
\setlength{\extrarowheight}{1.2pt}
\begin{adjustbox}{width=0.45\textwidth}
% \usepackage{multirow}
\begin{tabular}{ccccl}
\hline
\# \textbf{Bayesian layer}                           & \textbf{PSNR} $\uparrow$     & \textbf{SSIM} $\uparrow$     & \multicolumn{1}{c}{\textbf{GMSD} $\downarrow$}  & \multicolumn{1}{c}{\textbf{DSS} $\downarrow$} \\ \hline \hline
1           &   35.1125 & 0.9455 & 0.3022 &\textbf{0.0501}    \\
          2                   &   35.2456 & 0.9466 & 0.3145 &0.0598   \\ 
3 &  \textbf{35.4531} & \textbf{0.9476} & 0.3220  &  0.0698     \\
              4               &  35.0158& 0.9421 & \textbf{0.2985} & 0.0601    \\ \hline
\end{tabular}
\end{adjustbox}
\label{ablation bayesian layer}
\end{table}
\begin{figure*}[!t]
  \centering
  \subfigure[]{

  \begin{minipage}{0.3\textwidth}
  \centering
     \centerline{\includegraphics[width= 0.89\columnwidth]{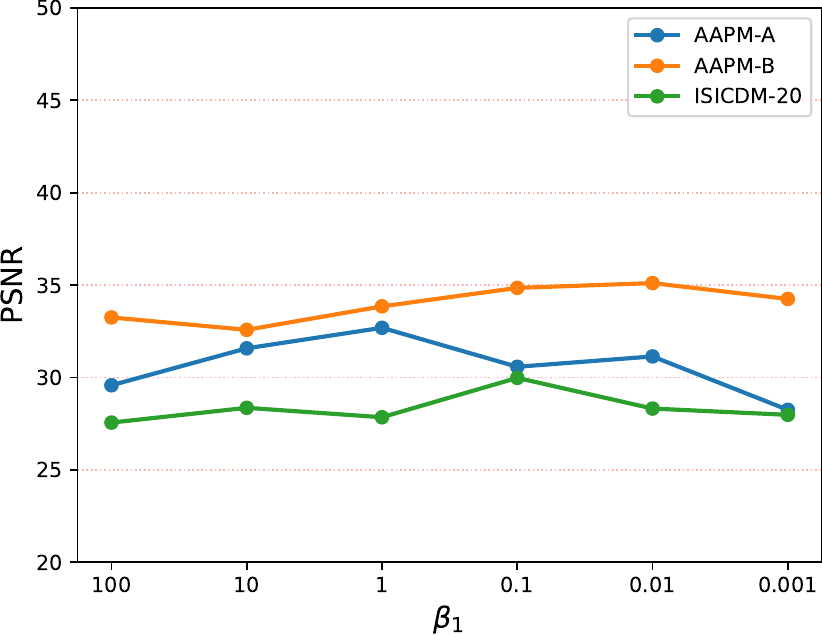}}
  \end{minipage}
  }
  \subfigure[]{

  \begin{minipage}{0.3\textwidth}
  \centering

     \centerline{\includegraphics[width= 0.89\columnwidth]{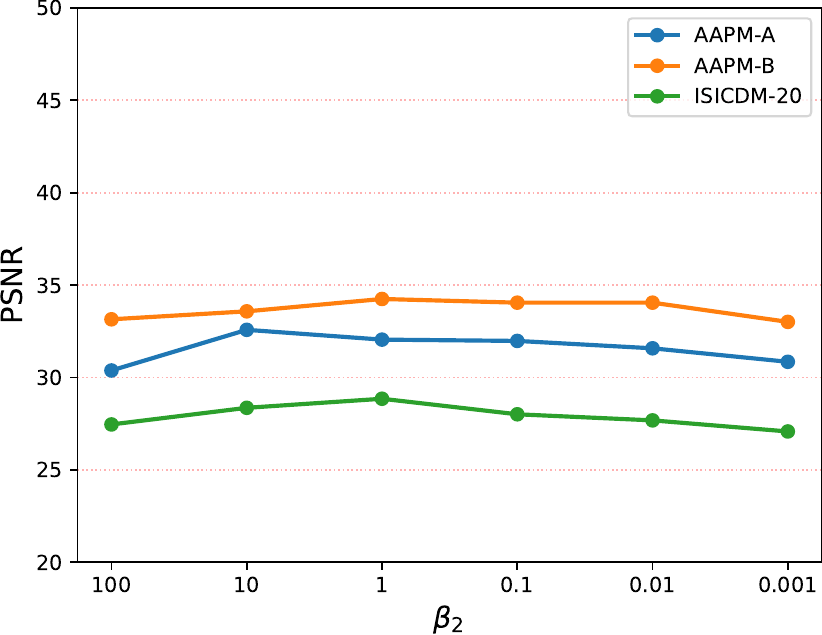}}
  \end{minipage}
  }
  \subfigure[]{

  \begin{minipage}{0.3\textwidth}
  \centering

     \centerline{\includegraphics[width= 0.89\columnwidth]{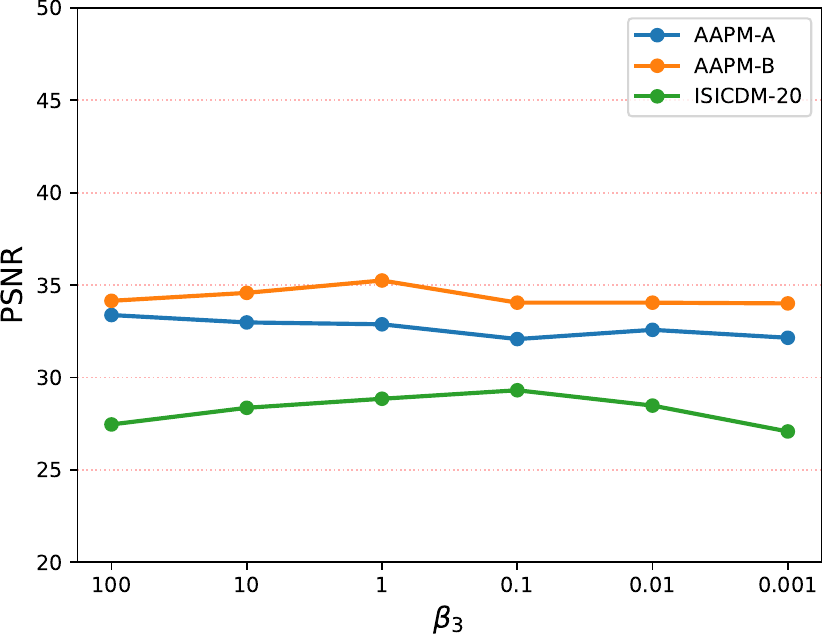}}
  \end{minipage}
  }
  \caption{Parameter sensitivity analysis by varying (a)$\beta_{1}$, (b)$\beta_{2}$, and (c) $\beta_{3}$. Each curve denotes the performance by considering the validate set on the source domain.}
  \label{parameter}
\end{figure*}

\begin{table*}[!t]
\centering
\caption{Subjective Quality Score (Mean ± SD) for Different Methods. The best and second-best performances are bolded and underlined, respectively.}
\setlength{\extrarowheight}{5pt}
\begin{adjustbox}{width=0.75\textwidth}
\begin{tabular}{cccccccccc}
\hline 
\multicolumn{1}{c}{}                & NDCT & LDCT & BM3D & Noise2Noise & CycleGAN & CCDnet & UDA & N2N-R & Ours \\ \hline
\multicolumn{1}{c}{Noise Reduction} &    -  &   -   &   \textbf{3.62}$\pm$0.23   &     3.02$\pm$0.12        &   2.98$\pm$0.23       &  3.21$\pm$0.34      &  3.25$\pm$0.34    &  3.05$\pm$0.41     &  \underline{3.55}$\pm$0.11    \\
Structure Preservation            &   -    &  -  &   2.57$\pm$0.54   &  3.36$\pm$0.43    &  3.28$\pm$0.36     &    3.25$\pm$0.23         &   3.40$\pm$0.33       &   \underline{3.56}$\pm$0.36     &   \textbf{3.78}$\pm$0.21     \\
Overall Clinical Usability          &   4.00    &  1.00  &   2.89$\pm$0.23   &  2.98$\pm$0.14    &  3.05$\pm$0.26     &    \underline{3.45}$\pm$0.25         &   3.41$\pm$0.36       &   3.39$\pm$0.35     &   \textbf{3.69}$\pm$0.29     \\ \hline
\end{tabular}
\end{adjustbox}
\label{double blind}
\end{table*}
\subsubsection{The comparison of inference time} We provide a comparison in terms of inference time in Table \ref{inference time}. As we can see, compared with other DL-based baseline approaches, e.g., Noise2Noise and CCDnet, our proposed method has a slight improvement in average inference time, which may be reasonable as our Bayesian framework needs extra MC sampling operation. In the future, it is feasible to consider a more efficient MC sampling strategy for less computation cost. However, our proposed method has higher computational efficiency compared with traditional methods and sinogram-domain counterparts.
\subsubsection{The comparison of different Bayesian layers}  We present an ablation study in terms of different numbers of Bayesian layers. The results can be found in Table \ref{ablation bayesian layer}. As we can see, the increase in Bayesian layer tasks marginal improvements. However, the computational cost will increase. Thus, one layer of the Bayesian layer may be sufficient for balancing the computational cost and mode performance.

\subsubsection{The analysis of hyperparameter selection} We provide parameter sensitivity analysis by varying $\beta_{1}$, $\beta_{2}$, and $\beta_{3}$ in Figure \ref{parameter}. As we can see, the Bayesian uncertainty alignment and sharpness-aware distribution alignment modules are relatively stable while
varying the parameters. Instead, the source domain-related hyperparameter, i.e., $\beta_{1}$ varies a lot, which may be reasonable as better cross-domain performance needs appropriate source knowledge to learn transferable representations. 

\subsection{Blind reader study}
We use 10 groups of CT slices on AAPM-A dataset for subjective low-dose CT reconstruction quality evaluation, where we score each reconstructed CT slice in
terms of noise reduction, structure preservation, and overall clinical usability by two
radiologists from Li Ka Shing Faculty of Medicine of Hong Kong University. Note that the score of overall
clinical usability on the five-point scale can effectively reflect the clinical task-related
assessment from radiologists’ perspective. The results of the double-blind study can be found in Table \ref{double blind}.  As we can see, our proposed method achieves the best or second-best performance among different methods. Note that our proposed method not only has the best overall clinical usability with the best structure preservation but also achieves second-best performance noise suppression. In all, our proposed method balances noise suppression and structural preservation well, leading to acceptable clinical usability.

\section{Discussion}
\textbf{The advantages of our proposed method over existing approaches.} 1) Compared with existing approaches, one of the most significant advantages is that our Bayesian neural networks-based framework can effectively correlate model uncertainty and cross-domain performance, which contributes to designing a BNN-based reconstruction framework in an uncertainty-guided manner. Instead, existing cross-domain LDCT image reconstruction approaches fail to do so due to their deterministic neural network-based frameworks. 2) More importantly, we devise a novel Bayesian uncertainty alignment in the latent space to reduce the epistemic gap between source and target domains, making it more likely to render well-reconstructed results on the target domain. 3) To address the inconsistent content information issue of distribution alignment, we carefully propose a sharpness-aware distribution alignment method in the image space to reduce the interference of content mismatch in adversarial distribution alignment.

\textbf{The rationality of MLV.} Other information can be represented by second-order information. For example, the luminance and contrast information using standard deviation in SSIM \cite{wang2004image} can measure image quality, which is similar to our proposed sharpness-aware measure based on maximum local variation (MLV). However, in the context of unsupervised domain adaptation, such a full-reference image quality assessment (IQA) approach like SSIM is not applicable as there are \textbf{no} ground-truth NDCT images on target domains. Instead, our proposed MLV-based sharpness-aware measure is a \textbf{no-reference IQA} method \cite{bahrami2014fast}, which can utilize the second-order information without the reference.
\textbf{Limitations and Future Works.} While the mean-field variational inference (MFVI) we employed is successful in providing an approximate posterior for Bayesian neural networks (BNNs), it has a potential limitation due to the fully factorized Gaussian assumption. This assumption restricts the expressiveness of the posterior distribution and may impede further enhancements in performance, especially on highly challenging CT reconstruction tasks \cite{cremer2018inference}. In the future,  it is feasible to investigate using more flexible and expressive approximate posteriors to overcome this limitation, such as normalizing flows. By incorporating such advanced techniques, we may improve the posterior approximation quality and potentially achieve better results on difficult medical image reconstruction tasks.  
\section{Conclusion}
This paper proposes a novel unsupervised domain adaptation method for LDCT image reconstruction using deep learning techniques. The method aims to address the issue of degraded reconstruction performance when deploying trained models to clinical CT imaging scenarios with variations in data acquisition. The proposed method leverages a probabilistic reconstruction framework and conducts joint discrepancy minimization between the source and target domains in both the latent and image spaces. In the latent space, a Bayesian uncertainty alignment technique reduces the gap between the two domains, improving the reconstruction of "unfamiliar" target domain data. In the image space, a sharpness-aware distribution alignment method ensures that the reconstructed images from the target domain have similar sharpness to normal-dose CT images from the source domain. Experimental evaluations demonstrate the effectiveness of the proposed method.
%%Harvard
\balance
\bibliographystyle{IEEEtran}

\begin{IEEEbiography}[{\includegraphics[width=1in,height=1.25in,clip,keepaspectratio]{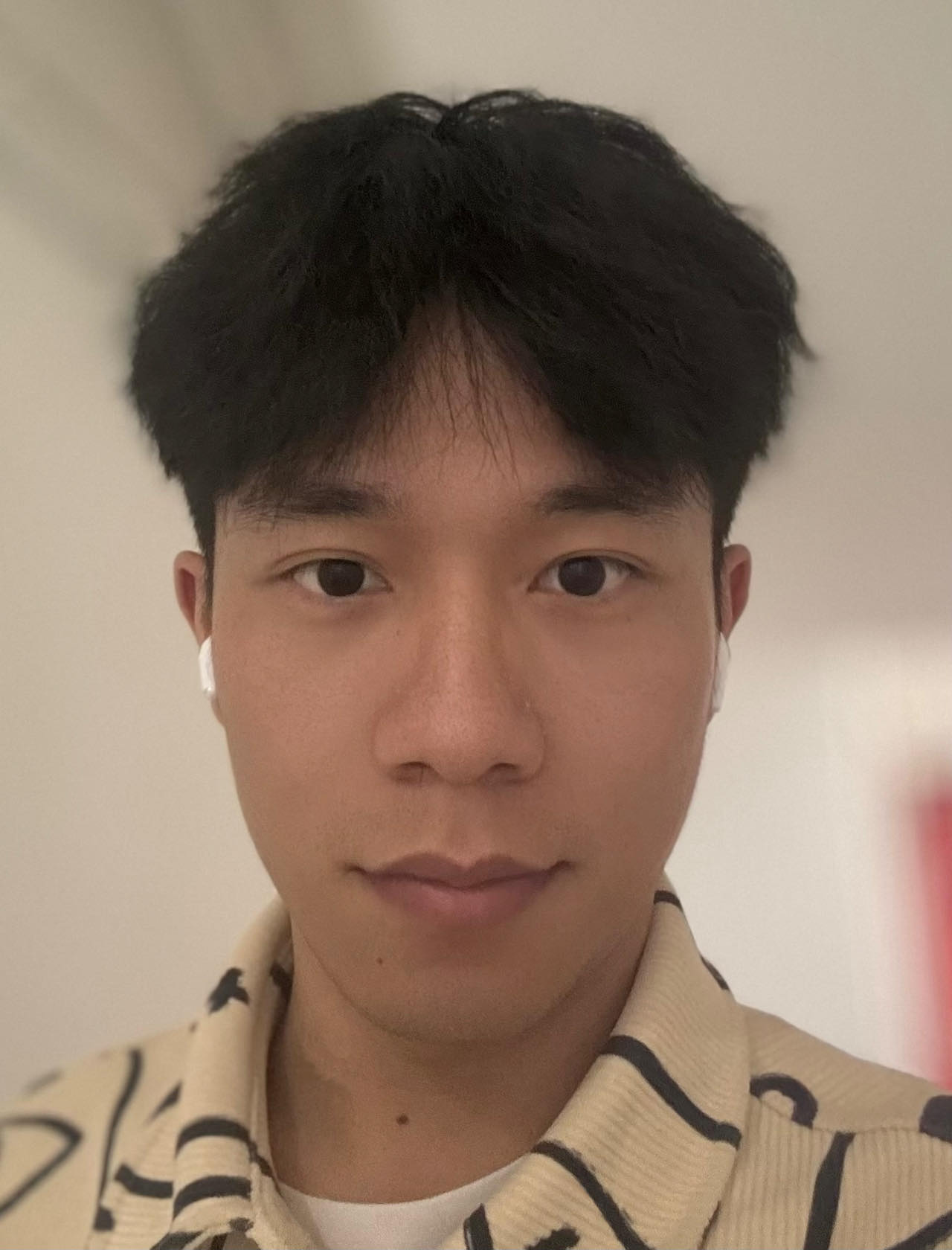}}]{Kecheng Chen}
received the B.S. degree and the M.S. degree from the Chengdu University of Technology and the University of Electronic Science and Technology of China, Sichuan, China, in 2019 and 2022, respectively. He is currently pursuing the Ph.D. degree with Department of Electrical Engineering, City University of Hong Kong, Hong Kong. His research interests include machine learning and transfer learning. 
\end{IEEEbiography}

\begin{IEEEbiography}[{\includegraphics[width=1in,height=1.25in,clip,keepaspectratio]{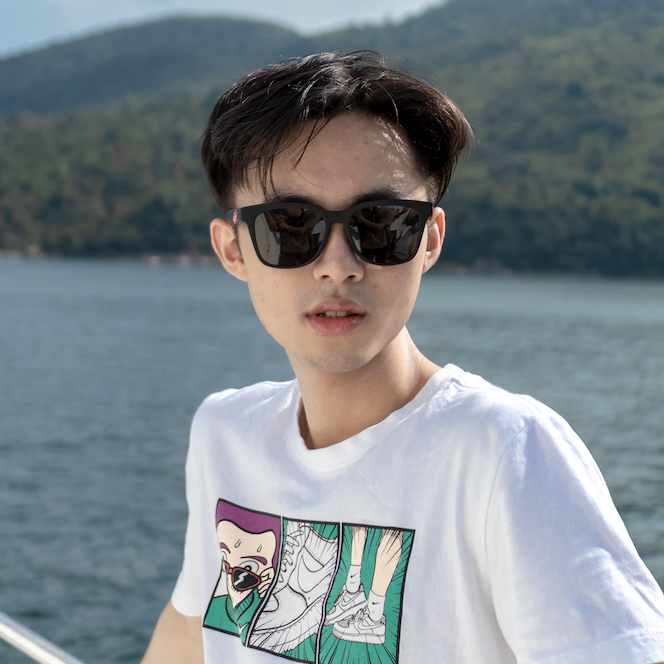}}]{Jie Liu}
received the BEng degree from the electronic science and technology, Zhejiang University, Zhejiang, in 2020. He is currently working toward the PhD degree with the Department of Electrical Engineering, City University of Hong Kong (CityU), Hong Kong. His research aims to develop novel machine learning algorithms for device-agnostic medical data processing and analysis. Technically, he focuses on distribution shift, open set learning, and partial label learning
\end{IEEEbiography}

\begin{IEEEbiography}[{\includegraphics[width=1in,height=1.25in,clip,keepaspectratio]{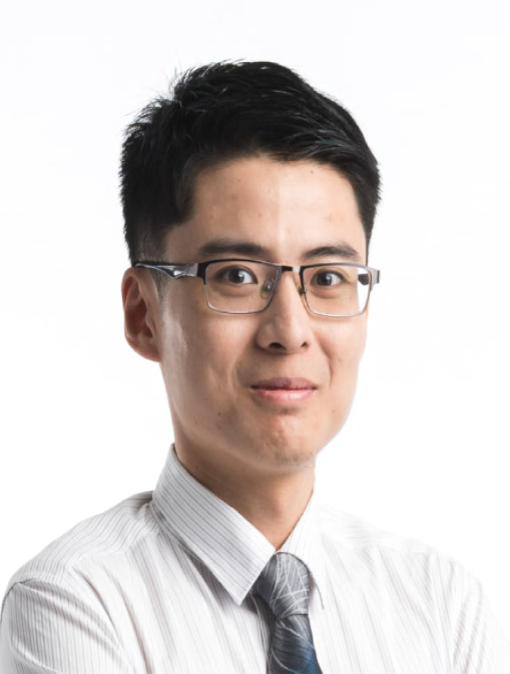}}]{Renjie Wan}
    received the B.Eng. degree from the University of Electronic Science and Technology of China in 2012 and the Ph.D. degree from Nanyang Technological University, Singapore, in 2019. He is currently an Assistant Professor with the Department of Computer Science, Hong Kong Baptist University, Hong Kong. He was a recipient of the Microsoft CRSF Award, the 2020 VCIP Best Paper Award, and the Wallenberg-NTU Presidential Postdoctoral Fellowship. He is the outstanding reviewer of the 2019 ICCV.
\end{IEEEbiography}

\begin{IEEEbiography}[{\includegraphics[width=1in,height=1.25in,clip,keepaspectratio]{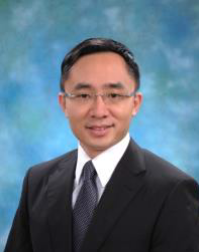}}]{Victor Ho-Fun  Lee}
is currently Chairperson and Clinical Associate
Professor of Department of Clinical Oncology, School of Clinical
Medicine, The University of Hong Kong. He was also Assistant Dean
(Assessment) of LKS Faculty of Medicine, The University of Hong
Kong between 2018 and 2024. He graduated in the University of
Hong Kong in 2002. After post-graduate residency training in clinical
oncology, he joined the Department of Clinical Oncology, Queen
Mary Hospital, The University of Hong Kong as Clinical Assistant
Professor in 2008. He obtained his fellowship in Royal College of
Radiologists in Clinical Oncology in 2010. Afterwards, he received
further specialist training in interstitial brachytherapy for head and
neck cancers and sarcoma in Institut Gustave Roussy in Paris, France
and novel radiation techniques like stereotactic radiosurgery and
stereotactic ablative radiotherapy in Stanford University USA. In
2013, he received further training on stereotactic body radiation
therapy for liver tumors at Princess Margaret Hospital, Toronto,
Canada. More recently in 2015 he was awarded HKCR 15A Traveling
Fellowship and pursued subspecialty training in image-guided
brachytherapy for cervical cancer and pediatric oncology. 
\end{IEEEbiography}

\begin{IEEEbiography}[{\includegraphics[width=1in,height=1.25in,clip,keepaspectratio]{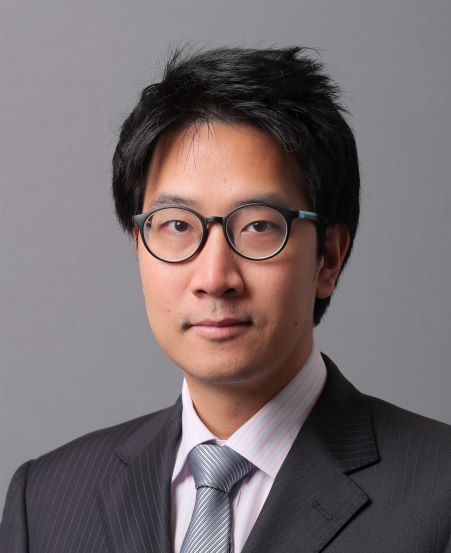}}]{Varut Vardhanabhuti}
completed his medical degree at Guy's, King's and St Thomas' School of Medicine in London, UK in 2005, and completed his radiology residency training at Imperial College London whilst also completing a PhD. Since then, he has been working as a Clinician Scientist at the University of Hong Kong currently serving as a Clinical Assistant Professor and co-director of the Medical AI Lab (MAIL) at HKUMed. His research's main focus is on the use of artificial intelligence in medical imaging to tackle complex age-related diseases such as atherosclerotic heart disease, cancers, and diabetes among others to predict outcomes or establish early biomarkers for subclinical disease. He also focuses on using deep learning for medical imaging reconstruction. The main focus is to accelerate the clinical adoption of these emerging technologies to benefit patients. He has published over 100 peer-reviewed publications (including high-impact journals such as Lancet Oncology, Lancet Digital Health, npjDigital medicine, JAMA open network, IEEE Transactions in Medical Imaging, Investigative Radiology, European Radiology, EJNMMI, among many others). He has more than 100 abstracts and presentations and has been invited to several local and international conferences as a keynote speaker.
\end{IEEEbiography}

\begin{IEEEbiography}[{\includegraphics[width=1in,height=1.25in,clip]{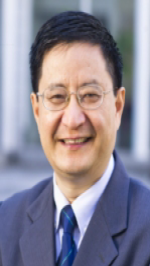}}]{Hong Yan (Fellow, IEEE)}
    received the Ph.D. degree from Yale University, New Haven, CT, USA, in 1989.,He was a Professor of Imaging Science with The University of Sydney, Camperdown, Australia. Currently, he is a Wong Chun Hong Professor of Data Engineering and a Chair Professor of Computer Engineering with the City University of Hong Kong, Hong Kong. He has authored or coauthored over 600 journal and conference papers in these areas. His research interests include image processing, pattern recognition, and bioinformatics.,Dr. Yan is an IAPR Fellow, a member of the European Academy of Sciences and Arts, and a Fellow of the US National Academy of Inventors. He received the 2016 Norbert Wiener Award for contributions to image and biomolecular pattern recognition techniques from the IEEE SMC Society.
\end{IEEEbiography}
\begin{IEEEbiography}[{\includegraphics[width=1in,height=1.25in,clip]{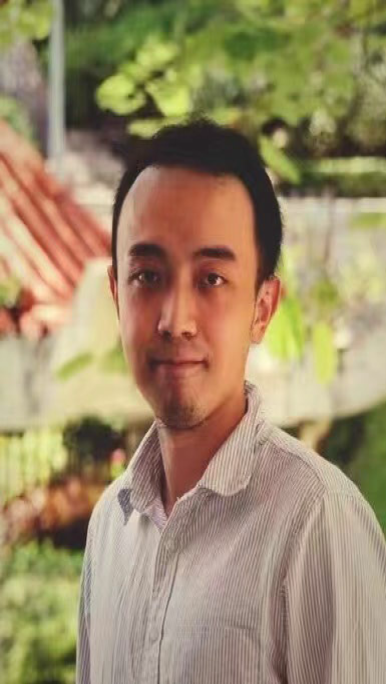}}]{Haoliang Li}
received the B.S. degree in communication engineering from the University of Electronic Science and Technology of China (UESTC), Chengdu, China, in 2013, and the Ph.D. degree from Nanyang Technological University (NTU), Singapore, in 2018. He is currently an Assistant Professor with the Department of Electrical Engineering, City University of Hong Kong. His research works appear in international journals/conferences, such as IEEE TRANSACTIONS ON PATTERN ANALYSIS AND MACHINE INTELLIGENCE (TPAMI), International Journal of Computer Vision (IJCV), IEEE TRANSACTIONS ON INFORMATION FORENSICS AND SECURITY (TIFS), NeurIPS, CVPR, and AAAI. His research interests include AI security, multimedia forensics, and transfer learning. He received the Wallenberg-NTU Presidential Postdoctoral Fellowship in 2019, the Doctoral Innovation Award in 2019,  the VCIP Best Paper Award in 2020, and ACM SIGSOFT distinguish paper award in 2022.
\end{IEEEbiography}

% or


\ifCLASSOPTIONcaptionsoff
  \newpage
\fi

\end{document}